\documentclass[pre,aps,showpacs,amsmath]{revtex4}
\usepackage{graphicx}
\begin{document}
\title{Solution of the spherically symmetric linear thermoviscoelastic problem in the inertia-free limit}
\author{Tage Christensen and Jeppe C. Dyre}
\affiliation{DNRF centre  ``Glass and Time,'' IMFUFA, Department of Sciences, Roskilde University, Postbox 260, DK-4000 Roskilde, Denmark}
\date{\today}

\newcommand{\iomt}{i\omega\tau}
\newcommand{\bu}{{\bf u}}
\newcommand{\br}{{\bf r}}
\newcommand{\bnul}{{\bf 0}}
\newcommand{\bnab}{{\bf\nabla}}
\newcommand{\pa}{\partial}
\newcommand{\rn}{\rho_0}
\newcommand{\dt}{\delta T}
\newcommand{\bv}{\beta_V}
\newcommand{\bj}{{\bf j}}
\newcommand{\lap}{\nabla^2}
\newcommand{\divu}{\bnab\cdot\bu}
\newcommand{\sk}{\sinh (k}
\newcommand{\ck}{\cosh (k}
\newcommand{\thma}{\mathbf{\tilde \Theta}}
\newcommand{\psima}{\mathbf{\tilde \Psi}}
\newcommand{\xma}{\mathbf{\tilde X}}
\newcommand{\ama}{\mathbf{\tilde A}}
\newcommand{\bma}{\mathbf{\tilde B}}
\newcommand{\cma}{\mathbf{\tilde C}}
\newcommand{\dma}{\mathbf{\tilde D}}
\newcommand{\mma}{\mathbf{\tilde M}}
\newcommand{\tma}{\mathbf{\tilde T}}
\newcommand{\gmma}{\mathbf{\tilde \Gamma}}
\newcommand{\yma}{\mathbf{\tilde Y}}
\newcommand{\hma}{\mathbf{\tilde H}}
\newcommand{\kma}{\mathbf{\tilde K}}
\newcommand{\lma}{\mathbf{\tilde L}}
\newcommand{\nma}{\mathbf{\tilde N}}
\newcommand{\pma}{\mathbf{\tilde P}}
\newcommand{\kk}{\textbf{K}}
\newcommand{\tth}{\textbf{T}^{\rm th}}
\newcommand{\rma}{\textbf {R}}
\newcommand{\tr}{\tilde r}
\newcommand{\tJ}{\tilde J}
\newcommand{\tit}{\tilde t}
\newcommand{\tf}{\tilde f}
\newcommand{\tu}{\tilde u}
\newcommand{\tA}{\tilde A}
\newcommand{\tP}{\tilde P}
\newcommand{\tal}{\tilde\alpha}
\newcommand{\dtpr}{\delta\tilde p_r}
\newcommand{\dtq}{\delta\tilde q}
\newcommand{\ttemp}{\tilde T}
\newcommand{\tg}{\tilde g}
\newcommand{\tc}{\tilde c}
\newcommand{\dtt}{\delta\tilde T}
\newcommand{\dtV}{\delta \tilde V}
\newcommand{\dtQ}{\delta \tilde Q}
\newcommand{\stress}{\underline{\underline{\sigma}}}
\newcommand{\strain}{\underline{\underline{\epsilon}}}
\newcommand{\dv}{\Delta V}

\begin{abstract}
The coupling between mechanical and thermal properties due to thermal expansion complicates the problem of measuring frequency-dependent thermoviscoelastic properties, in particular for highly viscous liquids. A simplification arises if there is spherical symmetry where -- as detailed in the present paper -- the thermoviscoelastic problem may be solved analytically in the inertia-free limit, i.e., the limit where the sample is much smaller than the wavelength of sound waves at the frequencies of interest. As for the one-dimensional thermoviscoelastic problem [Christensen {\it et al.}, Phys. Rev. E {\bf 75}, 041502 (2007)], the solution is conveniently formulated in terms of the so-called transfer matrix, which directly links to the boundary conditions that can be experimentally controlled. Once the transfer matrix has been calculated, it is fairly easy to deduce the equations describing various experimentally relevant special cases (boundary conditions that are adiabatic, isothermal, isochoric, etc.). In most situations the relevant frequency-dependent specific heat is the longitudinal specific heat, a quantity that is in between the isochoric and isobaric frequency-dependent specific heats.
\end{abstract}


\maketitle

\section{Introduction}

Linear thermoviscoelasticity is the well-established discipline dealing with the irreversible thermodynamics of slightly perturbed systems where mechanical and thermodynamic degrees of freedom couple to each other \cite{mei59,chr82,lan86,now86}. For glass-forming liquids cooled towards the calorimetric glass transition relaxation times become very long compared to phonon times, approaching and eventually exceeding seconds. For such ``ultraviscous'' liquids not only the mechanical moduli become complex and frequency dependent \cite{har76,win00,lak04}, but so do standard thermodynamic linear-response properties like the specific heat or the thermal expansion coefficient 
\cite{roe77,moy78,moy81,ang83,bir85,chr85,oxt86,jac90,chr94,hod94,nie96,chr97,chr98,boh00,kra02,chr07,chr95,ell07,gar07}.

Of the twelve basic complex, frequency-dependent thermodynamic linear-response coefficients only three are independent (see, e.g., Ref. \cite{ell07} and its references). If one assumes stochastic dynamics, which is believed to be realistic for viscous liquids on time scales much longer than phonon times, in fact only two thermoviscoelastic response functions are truly independent \cite{mei59,roe77,gup76,bai08}. No reliable measurements of a full set of (three) thermoviscoelastic response function appear to exist for any highly viscous liquid. Part of the reason for this may be the traditional focus in physics on phenomena on the molecular scale, but part of the problem most likely also comes from the considerable challenges associated with reliably measuring the thermoviscoelastic response functions. The problem is that the coupling between mechanics and thermodynamics caused by thermal expansion is nontrivial when the mechanical shear modulus is a significant fraction of the bulk modulus \cite{coupling_note}. This is true for solids in general, as well as for highly viscous liquids at frequencies of order the inverse relaxation time. For solids, however, because the isobaric and isochoric specific heats are almost identical, the problem is not serious. For ultraviscous liquids, on the other hand, the problem cannot be ignored. For such systems it was recently shown that the thermomechanical coupling implies that conventional methods fail to measure the isobaric, frequency-dependent specific heat \cite{chr07}. This is because truly isobaric conditions are difficult to establish experimentally due to the thermal expansion upon heating; in most experiments attempting to measure the frequency-dependent specific heat the stress tensor is not proportional to the unit tensor (i.e., there is not hydrostatic equilibrium), and shear stresses relax on the very time scale that one wishes to monitor.

The purpose of this paper is to establish the theoretical framework for measuring a complete set of thermoviscoelastic response functions utilizing spherical symmetry. The thermomechanical equations describing the coupling of mechanics to thermodynamics are well known since many years \cite{mei59,chr82}. We recently (with Olsen) presented the full analytic solution of the one-dimensional inertia-free case, i.e., where all motion is restricted to one direction and the wavelength of sound waves at the relevant frequencies is much larger than the sample size \cite{chr07}. The reader is referred to this paper as an introduction to the reasoning and techniques used in the present paper. Below, the spherically symmetric inertia-free case is treated, where the effective one-dimensional nature again allows the problem to be solved analytically. The solution is cumbersome, but once it has been arrived at a number of experimentally relevant special cases are fairly easy to calculate.

\begin{figure}
\includegraphics[width=6cm]{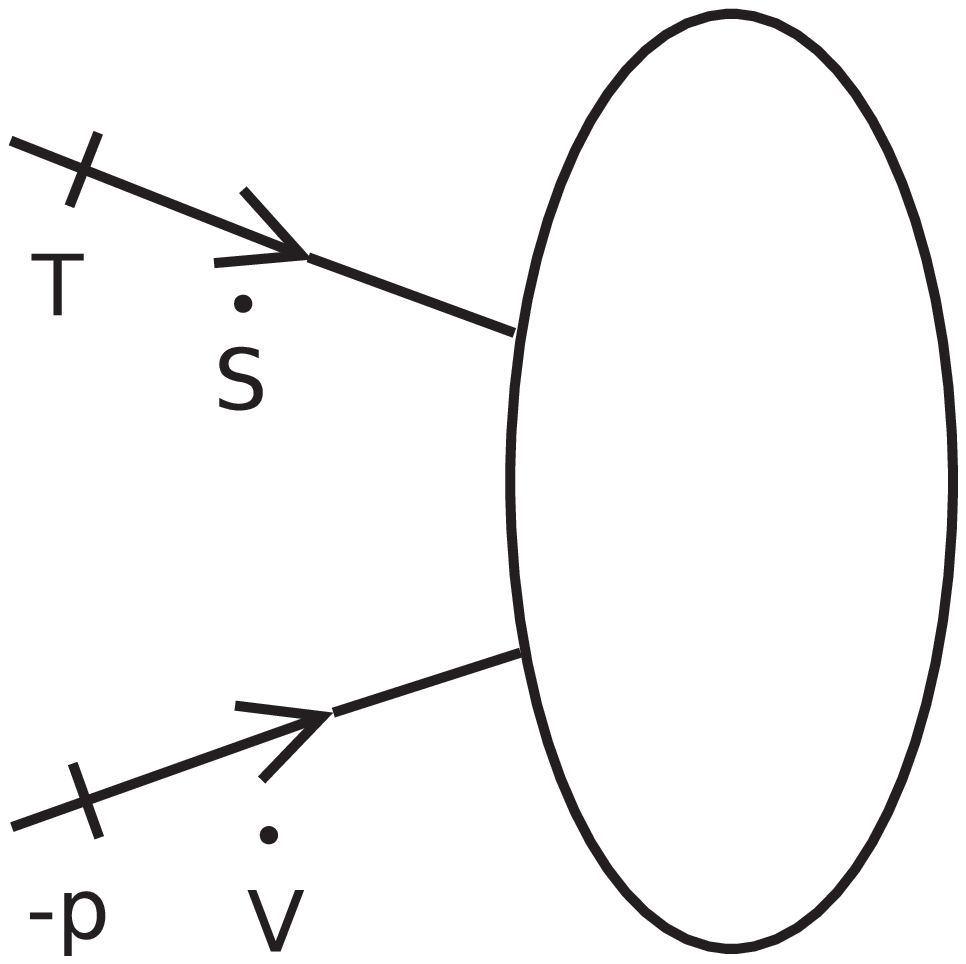}
\includegraphics[width=6cm]{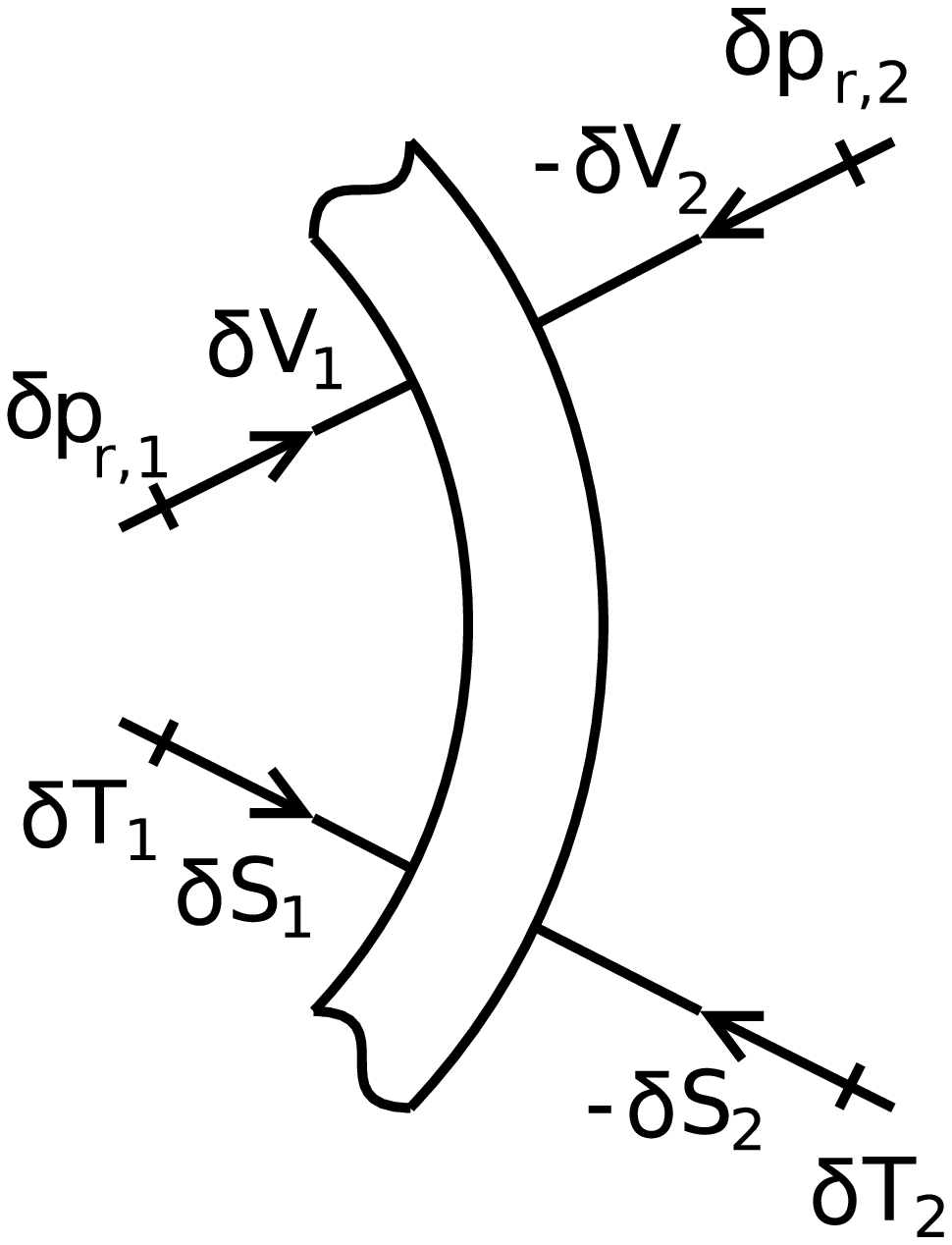}
\caption{Energy bond graphs \cite{ost73, mik93,pvc,kar06}, a useful tool for modeling linear thermoviscoelasticity. (a) The standard thermodynamic energy bonds, one thermal and one mechanical. In energy bond terminology temperature and negative pressure, $T$ and $-p$, are so-called effort variables analogous to voltage, and the entropy flux $\dot{S} = dS/dt$ and volume flux $\dot{V}= dV/dt$ are so-called flow variables analogous to electrical current. In quasi-equilibrium the product of an effort variable and its flow variable is the energy flux into the system from its surroundings; more generally it gives the flux of free energy - in the sense of available work - into the system. (b) Symbolic figure of the energy bonds of linear irreversible thermodynamics for the spherically symmetric situation. For the thermal energy bonds the efforts are the temperature variations $\delta T_1$ and  $\delta T_2$ at radius $r_1$ and $r_2$, respectively, and the flows are the entropy fluxes $\delta \dot{S}_1$ and  -$\delta \dot{S}_2$ into the system (fluxes are in the positive radial direction). For the mechanical energy bonds the flows are the volume fluxes $-\delta \dot{V}_1$ and  $\delta \dot{V}_2$ into the system, respectively, whereas pressure is now the ``radial pressure,'' $\delta p_{r,1}$ and  $\delta p_{r,2}$ respectively, defined in Eq. (\ref{b15}). The four energy bonds define eight variables. As shown in this paper, the fundamental physical equations provide four constraints among these eight variables. Thus there is a linear relation between the four ``outer'' and the four ``inner'' energy bond variables. This relationship is expressed in terms of the $4\times 4$ transfer matrix calculated below that provides all information needed to interpret any experimental situation characterized by specific boundary conditions.}
\end{figure}

The solution is formulated in terms of the so-called transfer matrix \cite{chr07,car59} that links infinitesimal variations at the boundaries for the following four quantities: entropy (or equivalently heat) input, volume displacement, pressure change, and temperature change. The transfer matrix -- not to be confused with the transfer matrix of statistical mechanical models -- is useful because it directly describes how the system interacts with its surroundings. These interactions are conveniently pictured and described via the energy bond technique \cite{ost73,mik93,pvc,kar06}. An energy bond is characterized by a {\it displacement} variable and an {\it effort} variable, generalizing the concepts of charge and voltage. The product of an effort and a differential displacement variable is a generalized work that gives the (free) energy transferred into the system from its surroundings.

In standard thermodynamics there are two energy bonds, one thermal and (if the shear modulus is negligible) one mechanical, see Fig. 1(a). The thermal energy bond is characterized by entropy $S$ as the displacement variable and temperature $T$ as the effort variable. For the mechanical energy bond the displacement is the volume $V$, and the effort is the negative pressure, $-p$. The time derivatives of displacement variables define {\it flow} variables, signaled by arrows in Fig. 1. The two energy bonds of Fig. 1(a) represent the fundamental thermodynamic identity 

\begin{equation}\label{fund}
dE
\,=\,TdS-pdV\,.
\end{equation}
In linear irreversible thermodynamics small perturbations in temperature and pressure are linear functionals of the volume and entropy flows, or vice versa. The temperature perturbation $d T=T-T_0$ is around a reference temperature $T_0$, and the pressure perturbation $dp=p-p_0$ around a reference pressure $p_0$. The supplied heat $dQ$ is related to the externally supplied entropy, $dS_{\rm ext}=dQ/T$. When the volume $V_0$ is so small that temperature and pressure can be assumed homogenous throughout the volume, the relaxation is described in the time domain by memory kernels of the form (where $\kappa_T$ is isothermal compressibility, $\alpha_p$ isobaric expansion coefficient, and $c_p$ isobaric specific heat per unit volume)

\begin{eqnarray}
dV(t)/V_0 &=& -\int_{-\infty}^t \kappa_T(t-t')dp(t')+\int_{-\infty}^t \alpha_p(t-t')dT(t')\,, \\
dS_{\rm ext}(t)/V_0 &=& -\int_{-\infty}^t \alpha_p(t-t')dp(t')+\int_{-\infty}^t \frac{1}{T_0}c_p(t-t')dT(t')\,.
\end{eqnarray}
Instead of Eq. (\ref{fund}) we now have

\begin{equation}\label{fund2}
dE
\,=\,TdS_{\rm ext}-pdV\,=\,  dT dS_{\rm ext}- dp dV + T_0 dS_{\rm ext} - p_0 dV.
\end{equation}
or

\begin{equation}\label{fund3}
d(E-T_0 S_{\rm ext} +p_0 V)\,=\,  dT dS_{\rm ext}- dp dV .
\end{equation}
The function $E-T_0 S_{\rm ext} +p_0 V$ is the imparted free energy, which is not a state function since dissipation in the system degrades this energy: During a cyclic process one has

\begin{equation}\label{dissip}
\oint d(E-T_0 S_{\rm ext} +p_0 V)=-T_0\oint dS_{\rm ext}=T_0 \Sigma\,,
\end{equation}
where $\Sigma$ is the entropy production in the system. Note that the entropy production is quadratic in the perturbations, implying that entropy is conserved to first order \cite{lan86}, a fact that is utilized below.

It follows from the above that the generic conjugated variables for a small system that relaxes mechanically and thermally with no shear forces are $(dT,dS_{\rm ext})$ and $(-dp,dV)$. From an experimental point of view, however, it is not convenient only to consider an infinitesimal volume, $V_0$. If one wishes to study relaxation on a time scale $\tau_{\rm exp}$, the volume $V_0$ may only be considered small if the heat diffusion time $\tau_D$ across the volume is much smaller than $\tau_{\rm exp}$. This is an unduly restriction that can only be coped with by analyzing the influence of heat diffusion on the response -- an important purpose of this paper. A further complication arises when the shear modulus becomes comparable to the bulk modulus, which is the case in the relaxation region of viscous liquids. In this case heat diffusion and mechanical stresses couple nontrivially. To keep the below discussion as simple as possible we look at the situation with highest symmetry, that of a sphere of inner radius $r_1$ and outer radius $r_2$, posing the question: What is the relation between the thermal and mechanical variables at the boundaries of the system? In order to take the shear forces properly into account it is shown below  that pressure must be replaced by the ``radial pressure,'' $\delta p_{r,1}$ at $r_1$ and  $\delta p_{r,2}$ at $r_2$. $\delta V_1$ is the volume swept by the surface at $r_1$ as a consequence of a radial, small displacement $u(r_1)$. Correspondingly, $\delta V_2$ is the volume swept at $r_2$ (both in the positive radial direction). The net volume change is $dV=-\delta V_1+\delta V_2$. The temperature perturbations at the two surfaces are denoted by $\delta T_1$ and $\delta T_2$, respectively, and the entropy fluxes in positive radial direction by $\delta S_1$ and $\delta S_2$, respectively. The net entropy influx is $dS_{\rm ext}=\delta S_1-\delta S_2$. Now Eq. (\ref{fund3}) becomes

\begin{equation}\label{fund4}
d(E-T_0 S_{\rm ext} +p_0 V)\,=\,  \delta T_1 \delta S_1- \delta T_2 \delta S_2 +\delta p_{r,1} \delta V_1-\delta p_{r,2} \delta V_2 .
\end{equation}
As illustrated in Fig. 1(b) this gives rise to four energy bonds, two referring to the outer radius $r_2$ and two to the inner radius $r_1$. 

The transfer matrix $\mathbf{T}(r_2,r_1)$ is by definition the $4\times 4$ matrix that links the (generally complex and frequency-dependent, see below) infinitesimal variations of the four energy bond variables at radius $r_1$ with the four energy bond variables at radius $r_2$. Switching notation such that $\delta p_{r,1}$ is denoted by $\delta p(r_1)$, etc, the transfer matrix is thus defined by 

\begin{equation}\label{transf44}
\begin{pmatrix}
\delta p(r_2)\\ 
\delta T(r_2)\\
\delta V(r_2)\\
\delta S(r_2)
\end{pmatrix}\,=\,
\mathbf{T}(r_2,r_1)
\begin{pmatrix}
\delta p(r_1)\\ 
\delta T(r_1)\\
\delta V(r_1)\\
\delta S(r_1)
\end{pmatrix}\,.
\end{equation}
The transfer matrix is related to the response matrix $\mathbf{\Gamma}(r_2,r_1)$ that by definition links the four effort variables to the four displacement variables:

\begin{equation}\label{resp44}
\begin{pmatrix}
\delta p(r_1)\\ 
\delta T(r_1)\\
\delta p(r_2)\\
\delta T(r_2)
\end{pmatrix}\,=\,
\mathbf{\Gamma}(r_2,r_1)
\begin{pmatrix}
\delta V(r_1)\\ 
\delta S(r_1)\\
-\delta V(r_2)\\
-\delta S(r_2)
\end{pmatrix}\,.
\end{equation}
From the fluctuation-dissipation theorem the response matrix is known to be symmetric, a fact that is explicitly confirmed below.

In the next section fundamentals are summarized. In Sec. III the full dynamic equations are formulated and brought into dimensionless form by scaling with complex units. In Sec. IV the equations are solved, and the transfer and response matrices are calculated. In Sec. V several experimentally relevant special cases are considered. Finally, Sec. VI gives a brief discussion.

\section{Definitions and constitutive relations}

The energy bond variables of standard thermodynamics give rise to a number of dc (i.e., static) linear-response coefficients as follows. If the variables of interest are those of the two thermodynamic energy bonds of Fig. 1(a), ($T\,, p\,, S\,, V$), there are altogether 24 thermodynamic coefficients of the form $(\partial a/\partial b)_c$ with $a,b,$ and $c$  chosen among $T, p, S$, and $V$ \cite{ber78,ell07}. These coefficients form 12 pairs that are trivially related by inversion [$(\partial a/\partial b)_c=1/(\partial b/\partial a) _c$, etc]. As is well known the 12 coefficients are not independent, but related by Maxwell relations. This leaves the following eight basic linear-response coefficients (where the specific heats here and throughout the paper are per unit volume): 

\begin{eqnarray}\label{thermo}
\mbox{Isochoric specific heat:}\,\,\,c_V &\equiv & \frac{T}{V} \Big(\frac{\partial S}{\partial T}\Big)_V  \\ 
\mbox{Isobaric  specific heat:}\,\,\, c_p &\equiv  &\frac{T}{V}   \Big(\frac{\partial S}{\partial T}\Big)_p\\
\mbox{Isothermal  compressibility:}\,\,\, \kappa_T &\equiv &-\frac{1}{V} \Big(\frac{\partial V}{\partial p}\Big)_T  \\ 
\mbox{Adiabatic compressibility:}\,\,\, \kappa_S &\equiv &-\frac{1}{V} \Big(\frac{\partial V}{\partial p}\Big)_S\\
\mbox{Isobaric expansion coefficient:}\,\,\, \alpha_p &\equiv & \frac{1}{V} \Big(\frac{\partial V}{\partial T}\Big)_p  = -\frac{1}{V}  \Big(\frac{\partial S}{\partial p}\Big)_T \label{iso}   \\
\mbox{Adiabatic contraction coefficient:}\,\,\, \alpha_S &\equiv &-\frac{1}{V} \Big(\frac{\partial V}{\partial T}\Big)_S = \frac{1}{V} \Big(\frac{\partial  S}{\partial p}\Big)_V  \label{ref_1} \\
\mbox{Isochoric pressure coefficient:}\,\,\, \beta_V &\equiv &\Big(\frac{\partial p}{\partial T}\Big)_V  = \Big(\frac{\partial S}{\partial V}\Big)_T   \\  \label{ref_2}
\mbox{Adiabatic pressure coefficient:}\,\,\, \beta_S &\equiv & \Big(\frac{\partial p}{\partial T}\Big)_S  = \Big(\frac{\partial S}{\partial   V}\Big)_p   \label{ref_3}
\end{eqnarray}
Some well-known relations between the thermodynamic coefficients are summarized in the Appendix for reference, where relations are simplified somewhat by letting the heat capacities be represented by the related variables $\zeta_V\equiv \frac{1}{V} ( \frac{\partial S}{\partial T} )_V$ and $\zeta_p\equiv \frac{1}{V} ( \frac{\partial S}{\partial T} )_p$. 

In systems with relaxing degrees of freedom the thermodynamic coefficients generally become complex and frequency dependent. Suppose, for instance, that the system is subjected to an infinitesimal periodic pressure variation with angular frequency $\omega$ described as $p(t)=p_0 +{\rm Re}[\delta p \exp(s t)]$, where $s=\pm i \omega$ depending on convention \cite{chr07} is the so-called Laplace frequency. The volume then varies periodically as $V(t)=V_0 +{\rm Re}[\delta V \exp(s t)]$, where both $\delta p$ and $\delta V$ are generally complex and frequency dependent. If this takes place at constant temperature, the complex frequency-dependent isothermal compressibility is defined via $\kappa_T\equiv -\delta V / (\delta p V_0)$, where $V_0$ is the average volume. According to a basic theorem of linear irreversible thermodynamics, the Maxwell relations among the dc linear-response quantities in Eqs. (6)-(9) translate into Onsager relations for the frequency-dependent coefficients, which reflect time reversibility. This is a special case of the so-called correspondence principle \cite{mei59,chr82,win00}: Any (dc) thermodynamic relation or equation involving linear thermodynamic and/or mechanical quantities applies unchanged when constitutive properties are replaced by the corresponding complex, frequency-dependent quantities. These frequency domain functions are related to the corresponding memory kernels like, e.g.,
\begin{equation}
 \kappa_T(s)=s \int_0^\infty \kappa_T(t) e^{-st}dt\,,
\end{equation}
where we adhere to the ordinary (sloppy) notation of physics that it is to be read out of the context whether $\kappa_T$ refers to time or frequency domain function. In the following we stay in the frequency domain, though.

Thermoviscoelasticity describes the coupling between thermal and mechanical deviations from equilibrium. This paper deals with the linear case that is well understood as regards fundamentals. Linearity means that the system is assumed to be infinitesimally close to equilibrium. Deviations from equilibrium are quantified in terms of the infinitesimal displacement field $\bu=\bu(\br,t)$, temperature variation field $\dt(\br,t)=T(\br,t)-T_0$, etc. In this approximation the radial heat displacement is given by $\delta Q(r)=\delta S(r)/T_0$;  we switch to $Q$ simply because heat capacity traditionally is defined via this quantity.  

The isothermal bulk modulus $K_T$ (inverse isothermal compressibility) is defined by

\begin{equation}\label{KT_def}
K_T\,\equiv\,-V\left(\frac{\pa p}{\pa V}\right)_T
\,=\,\frac{1}{\kappa_T}\,.
\end{equation}
If $\pa_i$ is the derivative with respect to the i'th spatial coordinate $x_i$ [where $(x,y,z)=(x_1,x_2,x_3)$] and $u_i$ is the i'th component of the infinitesimal displacement vector $\bu$, the strain tensor $\underline{\underline{\epsilon}}=\epsilon_{ij}$ is defined \cite{lan86} by 

\begin{equation}\label{stran_def}
\epsilon_{ij}=\frac{\pa_iu_j+\pa_ju_i}{2}\,.
\end{equation}
The relative volume change is given by the trace of the strain tensor:

\begin{equation}\label{dv_over_v}
\frac{\delta V}{V}\,=\,
{\rm tr}(\strain)=\divu\,.
\end{equation}
Denoting the stress tensor by $\underline{\underline{\sigma}}=\sigma_{ij}$, the shear modulus $G$ is defined \cite{lan86} as follows (note that the isothermal and adiabatic shear moduli are always identical):

\begin{equation}\label{b14}
\sigma_{ij}-\frac{1}{3}{\rm tr}(\stress)\delta_{ij}=2G\left(\epsilon_{ij}-\frac{1}{3}{\rm tr}(\strain)\delta_{ij}\right)\,.
\end{equation}
If there are both infinitesimal displacements and infinitesimal spatial temperature variations, which is generalized into the so-called Duhamel-Neumann relation \cite{lan86,now86} that is the following constitutive relation linking mechanical and thermodynamic properties:

\begin{equation}\label{duhamel}
\sigma_{ij}\,=\,K_T{\rm tr}(\strain)\delta_{ij}+2G\left(\epsilon_{ij}-\frac{1}{3}{\rm tr}(\strain)\delta_{ij}\right)-\bv\dt\delta_{ij}\,.
\end{equation}
For relaxing systems $K_T$ and $G$ are complex and frequency dependent, and for periodically varying boundary conditions both strain and stress tensors are generally complex. By the correspondence principle the Duhamel-Neumann relation applies also for the frequency-dependent case.

The isothermal longitudinal modulus $M_T$ is defined \cite{lan86} by

\begin{equation}\label{M_def}
M_T\,=\,
K_T+\frac{4}{3}G\,.
\end{equation}
Similarly, the adiabatic longitudinal modulus $M_S$ is defined by $M_S=K_S+(4/3)G$ where $K_S=1/\kappa_S$ is the adiabatic bulk modulus. As shown in detail in Ref. \cite{chr07} (but implicit \cite{bem} already in Ref. \cite{lan86}), a useful quantity for the description of one-dimensional thermoviscoelasticity is the ``longitudinal specific heat'' defined by

\begin{equation}\label{long_spec}
c_l\,\equiv\,
c_V\,+\,T_0\frac{\bv^2}{M_T}\,.
\end{equation}
An important result of the present paper is that the longitudinal specific heat also plays a significant role for the spherically symmetric case. Again, $c_l$ is frequency dependent for relaxing systems. A useful identity \cite{chr07} follows from Eqs. (A.1) and (A.7):

\begin{equation}\label{clid}
c_l\,=\,c_V\,\frac{M_S}{M_T}=\,c_V\,\frac{K_S+\frac{4}{3}G}{K_T+\frac{4}{3}G}\,.
\end{equation}
From this one finds that $c_l$ may be interpreted as a generally complex convex combination of $c_p$ and $c_V$:

\begin{equation}\label{cl_conv}
c_l\,=\,
\frac{3K_T}{3K_T+4G}\,c_p\,+\,\frac{4G}{3K_T+4G}\,c_V\,.
\end{equation}
This shows that the longitudinal heat capacity is effectively in between the isobaric and isochoric heat capacities. In analogy to the standard abbreviation 

\begin{equation}\label{cp_cv_rel}
\gamma\,\equiv\,\frac{c_p}{c_V}\,,
\end{equation}
we define $\gamma_l$, a quantity that is also generally complex and frequency dependent, as follows \cite{chr07}:

\begin{equation}\label{cl_cv_rel}
\gamma_l\,\equiv\,\frac{c_l}{c_V}\,.
\end{equation}

If the heat-current density is denoted by $\bj$, the heat conductivity $\lambda$ is defined via Fourier's law,

\begin{equation}\label{jq_def}
\bj=-\lambda\bnab\dt\,.
\end{equation}
The heat conductivity is generally assumed to be frequency independent, an assumption that was recently confirmed \cite{min01,ben03}. The below solution, however, applies also if $\lambda$ were to depend on frequency.

\section{The coupled equations for the temperature and displacement fields}

\subsection{The frequency-independent case}

As mentioned, entropy conservation always applies to first order, a fact that in Ref. \cite{lan86} is referred to as ``the equation of continuity for heat.'' In terms of the infinitesimal displacement field $\bu(\br,t)$ and the infinitesimal temperature field $\dt(\br,t)$, for a volume element $dV$ at average mass density $\rho_0$, the basic thermoviscoelastic equations of motion reflecting entropy conservation and Newton's second law, $\rho_0dV d^2 u_i/dt^2=dV\sum_j\partial_j\sigma_{ij}$  \cite{chr07,chr82,lan86,now86} are (utilizing Eq. (\ref{duhamel})):

\begin{equation}\label{N_2}
\rn\frac{\pa^2\bu}{\pa t^2}\,=\,
M_T\bnab(\divu)-G\bnab\times(\bnab\times\bu)-\bv\bnab\dt
\end{equation}
and

\begin{equation}\label{thermo_eq}
c_V\frac{\pa\,\dt}{\pa t}+T_0 \bv \frac{\pa}{\pa t}\left(\divu\right)
\,=\,\lambda\lap\dt\,.
\end{equation}
We shall only be concerned with the inertia-free limit, i.e., the limit where the sample is much smaller than the wavelength of sound waves at the frequencies of interest. In this limit the acceleration term is negligible, thus

\begin{eqnarray}
M_T\bnab(\divu)-G\bnab\times(\bnab\times\bu)-\bv\bnab\dt\,&=&\,0\label{b1}\\
c_V\frac{\pa\dt}{\pa t}+T_0 \bv\frac{\pa}{\pa t}\left(\divu\right)&\,=\,&\lambda\lap\dt\label{b2}\,.
\end{eqnarray}
Before proceeding we note that the identity $\bv\,=\,\alpha_p\, K_T$ [Eq. (A5)] implies that if there is no thermal expansion upon heating, then $\bv=0$. In this case the two equations decouple and reduce to the ordinary elastic equation of motion in the inertia-free limit and the heat-conduction equation, respectively. Thus the coupling between mechanics and thermodynamics arises only when the thermal expansion coefficient is nonzero, as is indeed intuitively obvious.

The pressure variation $\delta p$ is defined by (assuming here and henceforth that the average stress tensor is zero)

\begin{equation}\label{b3}
\delta p=-\frac{1}{3} {\rm tr}\left(\underline{\underline{\sigma}}\right)\,.
\end{equation}
This definition applies also for nonhydrostatic conditions, i.e., when $\underline{\underline{\sigma}}$ is not proportional to the unit tensor. When the specific heat or the expansion coefficient is termed isobaric, it refers to a situation where the trace of the stress tensor is constant, not necessarily its individual diagonal components. The relative volume change of Eq. (\ref{dv_over_v}) is related to changes in pressure, $\delta p$, and temperature, $\dt$, by the following equation (see, e.g., Ref. \cite{chr07})

\begin{equation}\label{b4}
 \divu=-\frac{1}{K_T}\delta p + \alpha_p \dt\,.
\end{equation}
Using Eqs. (\ref{b4}) and (A1) one finds that under isobaric conditions $T_0 \bv \divu=T_0 \bv\alpha_p \dt=\left(c_p-c_V \right) \dt$. Thus Eq. (\ref{b2}) becomes the standard heat-diffusion equation

\begin{equation}\label{b7}
\frac{\pa\dt}{\pa t}=D_p\lap\dt\,,
\end{equation}
where the heat-diffusion constant $D_p$ involves the isobaric specific heat, $D_p=\lambda/c_p$. In general, isobaric conditions do not apply and the full coupled system Eqs. (\ref{b1}) and (\ref{b2}) must be solved.

If $\bnab\times\bu=\bf 0$, Eq. (\ref{b1}) simplifies considerably. This applies for the spherically symmetric case, $\bu=u(r) \bf \hat{r}$, where Eq. (\ref{b1}) reduces to

\begin{equation}\label{b8}
\bnab(M_T\divu-\bv \dt)=0\,.
\end{equation}
This is immediately integrated to

\begin{equation}\label{b9}
\divu=\frac{\bv}{M_T}\dt+a_1\,.
\end{equation}
Here $a_1$ is an integration constant that is a function of time determined by the boundary conditions. Substituting Eq. (\ref{b9}) into Eq. (\ref{b2}) yields

\begin{equation}\label{b11}
c_l\frac{\pa\dt}{\pa t}+T_0 \bv \frac{\pa a_1}{\pa t}=\lambda\lap\dt\,.
\end{equation}

\subsection{The case of spherical symmetry and periodically varying fields}

We now assume that all fields depend only on radius $r$ and that their time dependence is harmonic, i.e., $\propto \exp(st)$ [$s=\pm i\omega$]. Henceforth $u$ and $\dt$ refer to the complex frequency-dependent amplitudes of the infinitesimal radial displacement and temperature fields, respectively. All constitutive quantities are generally complex functions of the Laplace frequency $s$.

With these assumptions, if differentiation with respect to $r$ is denoted by a prime, using the identity $\divu=r^{-2}(r^2u)'$ Eqs. (\ref{b9}) and (\ref{b11}) become

\begin{equation}\label{b12}
r^{-2}(r^2u)'=\frac{\bv}{M_T}\dt+a_1
\end{equation}
and

\begin{equation}\label{b13}
c_l s \dt+ T_0 \bv s a_1=\lambda r^{-2}(r^2 \dt')'\,.
\end{equation}
Throughout the paper it is important to remember that not only the field amplitudes $u$ and $\dt$ are generally complex functions of $s$, but so are all constitutive properties like $\bv$, $M_T$, etc. For simplicity of notation, however, the frequency dependence will usually not be explicitly indicated.

We need two further equations for two auxiliary fields. One is the ''radial pressure,`` 

\begin{equation}\label{rad_pres}
\delta p_r \,\equiv\,-\sigma_{rr}\,,
\end{equation}
i.e., the normal force per unit area in the negative radial direction exerted on a spherical surface of radius $r$ from the outside. The radial pressure is generally different from the pressure $\delta p$ of Eq. (\ref{b3}); the relation between the two follows from the definition of the shear modulus Eq. (\ref{b14}):

\begin{equation}\label{b15}
 \delta p_r= \delta p - 2 G\left(\epsilon_{rr}-\frac{1}{3}\divu\right)\,.
\end{equation}
If $G=0$ -- in particular for any liquid at zero frequency -- radial pressure equals pressure. Since $\epsilon_{rr}=u'$ where $\epsilon_{rr}$ is the $rr$'th component of the strain tensor in spherical coordinates \cite{lan86}, Eq. (\ref{b15}) and the Duhamel-Neumann relation Eq. (\ref{duhamel}) imply

\begin{equation}\label{b16}
\delta p_r=-(K_T+\frac{4}{3}G)u'-(K_T-\frac{2}{3}G)\frac{2}{r}u+K_T\alpha_p \dt\,.
\end{equation}
The other auxiliary field to be introduced is the time-integrated heat-current density $\delta q$. Invoking Fourier's law of heat conduction Eq. (\ref{jq_def}) we get

\begin{equation}\label{b17}
 \delta q
\,=\,\frac{1}{s}j
\,=\,-\frac{\lambda}{s}\dt'\,.
\end{equation}

\subsection{Scaling to dimensionless variables}

The solution of the thermoviscoelastic equations in spherical symmetry is extremely involved when expressed in terms of the original variables. We solve the equations below by proceeding in analogy to the solution of the one-dimensional problem detailed in Ref. \cite{chr07} where considerable simplification was obtained by scaling with suitable variables. One is free to scale by a characteristic length, time, temperature, and mass, or other combinations of these four fundamental dimensions. We choose pressure instead of mass as the fourth fundamental dimension. There is no mathematical requirement that scaling variables must be real; in the frequency domain they may well be complex and frequency dependent. 

The scaling units are chosen as follows. The unit of time is the inverse complex Laplace frequency, $s^{-1}$. The unit of length is the (complex and frequency dependent) heat-diffusion length

\begin{equation}\label{ld_def}
l_D\,\equiv\,\sqrt{\frac{\lambda}{c_ls}}\,.
\end{equation}
It is occasionally convenient to use the (complex and frequency dependent) wave number $k\equiv 1/l_D$,

\begin{equation}\label{k_def}
k
\,=\,\sqrt{\frac{c_ls}{\lambda}}\,.
\end{equation}
The unit of temperature is the average temperature $T_0$, and the unit of pressure is the (complex and frequency dependent) isothermal bulk modulus $K_T$. With these units we define the following dimensionless variables

\begin{eqnarray}\label{b19tob25}
\tit\,&\equiv&\,st\\
\tr\,&\equiv&\,r/l_D\\
\tu\,&\equiv&\,u/l_D\\
\dtt\,&\equiv&\,\dt/T_0\\
\dtpr\,&\equiv&\,\delta p_r/K_T\\
\dtq\,&\equiv&\,\delta q/(K_Tl_D)\\
\tc\,&\equiv&\,T_0c_l/K_T\\
\tg\,&\equiv&\,4G/(3K_T)\\
\tal\,&\equiv&\,T_0\alpha_p\,.
\end{eqnarray}
It should once again be emphasized that all thermodynamic and mechanical constitutive quantities are generally complex and frequency dependent, although for simplicity of notation this fact is rarely explicitly indicated below.

The scaled problem involves only the three independent frequency-dependent response functions: $\tc$, $\tg$, and $\tal$. Nevertheless, it is convenient to introduce two further dimensionless frequency-dependent parameters, $\gamma$ and $\gamma_l$ of Eqs. (\ref{cp_cv_rel}) and (\ref{cl_cv_rel}), which are functions of $\tc$, $\tg$, and $\tal$. Before proceeding we note that via the above definitions, Eqs. (A.1), (A.4), and (A.5), the parameters $\gamma$ and $\gamma_l$ obey

\begin{eqnarray}\label{b26a}
\frac{1}{\gamma_l}
\,&=&\,1-\frac{\tal^2}{\tc(1+\tg)}\,,\\
\label{b27a}
\frac{\gamma}{\gamma_l}
\,&=&\,1+\frac{\tal^2\tg}{\tc(1+\tg)}\,.
\end{eqnarray}
These equations imply the following identity that turns out to be useful,

\begin{equation}\label{g_gamma_id}
\gamma+\tg \,=\,
\gamma_l\left(1+\tg\right)\,.
\end{equation}
When rewritten in terms of the above dimensionless variables, Eqs. (\ref{b13}), (\ref{b12}), (\ref{b16}), and (\ref{b17}) become (where prime now implies differentiation with respect to $\tr$)

\begin{eqnarray}\label{b26}
\tr^{-2}(\tr^2 \dtt')'&=&\dtt+\frac{\tal}{\tc}a_1\,,\\
\label{b27}
\tr^{-2}(\tr^2 \tu)'&=&\frac{\tal}{1+\tg}\dtt+a_1\,,\\
\label{b28}
\dtpr&=&-(1+\tg)\tu'-(2-\tg)\tr^{-1}\tu+\tal\dtt\,,\\
\label{b29}
\dtq&=&-\tc\dtt'\,.
\end{eqnarray}
These are the fundamental dimensionless thermoviscoelastic equations of spherical symmetry to be solved.

\section{Solution in terms of the transfer matrix}

Equation (\ref{b27}) leads to

\begin{equation}
\tu'=\frac{\tal}{1+\tg}\dtt+a_1-2\tr^{-1}\tu\,,
\end{equation}
by which Eq. (\ref{b28}) is simplified as follows

\begin{equation}\label{31}
\dtpr=-(1+\tg)a_1+3\tg \tr^{-1}\tu\,.
\end{equation}
Equation (\ref{b26}) implies $(2/\tr)\dtt'+\dtt''=\dtt+(\tal/\tc)a_1$, which is readily solved by substituting $\dtt=f(\tr)/\tr$. The solution is

\begin{equation}
\dtt=-\frac{\tal}{\tc}\, a_1+\tr^{-1} e^{\tr} \, a_3+\tr^{-1} e^{-\tr} \, a_4\,,
\end{equation}
where $a_3$ and $a_4$ are integration constants. Substituting this into Eq. (\ref{b27}) and solving for $\tu$ gives (where $a_2$ is a further integration constant)

\begin{equation}
\tu=\frac{1}{3}\left(1-\frac{\tal^2}{(1+\tg)\tc}\right)\tr \, a_1+\tr^{-2}\, a_2+ \frac{\tal}{1+\tg}\Big((\tr^{-1}-\tr^{-2})e^{\tr} a_3 - (\tr^{-1}+\tr^{-2})e^{-\tr} \, a_4\Big)\,.
\end{equation}
Now Eq. (\ref{31}) becomes

\begin{equation}
\dtpr=-\left(1+\frac{\tg}{1+\tg}\frac{\tal^2}{\tc}\right)\, a_1+3\tg \tr^{-3} \, a_2-3\frac{\tg}{1+\tg}\tal \tr^{-3}\Big((1-\tr) e^{\tr}\, a_3+(1+\tr) e^{-\tr}\, a_4\Big)\,.
\end{equation}
Finally we have

\begin{equation}
\dtq=\tc \tr^{-2}[(1-\tr)e^{\tr} \, a_3 +(1+\tr)e^{-\tr} \, a_4]\,.
\end{equation}
The solution is summarized in the form

\begin{equation}\label{b30}
\begin{pmatrix}
\dtpr(\tr)\\ 
\dtt(\tr)\\
\dtV(\tr)\\
\dtQ(\tr)
\end{pmatrix}\,=\,
\mma(\tr)
\begin{pmatrix}
a_1\\ a_2 \\ a_3 \\a_4
\end{pmatrix}
\,.
\end{equation}
Here, if $\delta Q(r)$ is the time-integrated heat current through the spherical surface of radius $r$, we have introduced 

\begin{eqnarray}\label{dvtildedef}
\dtV\, &\equiv &\,\tr^2 \tu\,\,=\frac{\delta V}{4\pi l_D^3}\,,\\
\dtQ  &\equiv &\,\tr^2 \dtq\,\, =\frac{\delta Q(r)}{4\pi K_Tl_D^3}\,.
\end{eqnarray}
The matrix $\mma(\tr)$ is given by

\begin{equation}\label{b31}
\mma(\tr)\,=\,
\begin{pmatrix}
-(1+\frac{\tg}{1+\tg}\frac{\tal^2}{\tc}) & 3 \tg \tr^{-3} & -3\frac{\tg}{1+\tg}\tal \tr^{-3}(1-\tr) e^{\tr} & -3\frac{\tg}{1+\tg}\tal \tr^{-3}(1+\tr) e^{-\tr}\\
-\frac{\tal}{\tc} & 0 & \tr^{-1}e^{\tr} & \tr^{-1}e^{-\tr}\\
\frac{1}{3}(1-\frac{\tal^2}{\tc(1+\tg)})\tr^3 & 1 & -\frac{\tal }{1+\tg} (1-\tr)e^{\tr}& -\frac{\tal }{1+\tg} (1+\tr)e^{-\tr}\\
0 & 0 & \tc (1-\tr)e^{\tr} & \tc (1+\tr) e^{-\tr}
\end{pmatrix}\,.
\end{equation}
Defining the transfer matrix by

\begin{equation}\label{b32}
 \tma(\tr_2,\tr_1)\,\equiv\,\mma(\tr_2)\mma^{-1}(\tr_1)\,,
\end{equation}
allows one to express the fields at $\tr_2$ in terms of those at $\tr_1$ as follows

\begin{equation}\label{b33}
\begin{pmatrix}
\dtpr(\tr_2)\\ 
\dtt(\tr_2)\\
\dtV(\tr_2)\\
\dtQ(\tr_2)
\end{pmatrix}\,=\,
\tma(\tr_2,\tr_1)
\begin{pmatrix}
\dtpr(\tr_1)\\ 
\dtt(\tr_1)\\
\dtV(\tr_1)\\
\dtQ(\tr_1)
\end{pmatrix}\,.
\end{equation}
In this way the inner and outer boundary conditions of Fig. 1(b) are linked by a matrix containing all relevant information about the physics of the system. The transfer matrix is $4\times 4$, reflecting the fact that out of the eight variables of the four energy bonds of Fig. 1(b), four may be externally controlled. Via the transfer matrix the remaining four are determined by the four basic equations (\ref{b1}) and (\ref{b2}). Once the transfer matrix has been calculated, various experimentally relevant special cases may be worked with modest efforts (Sec. V).

By explicit calculation, e.g., via a computer program, one finds that $\tma(\tr_2 ,\tr_1)$ has the form

\begin{equation}\label{b34}
\tma(\tr_2 ,\tr_1)=\tma^0(\tr_2 ,\tr_1)+\tma^-(\tr_2 ,\tr_1)e^{-(\tr_2-\tr_1)}+\tma^+(\tr_2 ,\tr_1)e^{\tr_2-\tr_1}
\end{equation}
where

\begin{equation}\label{b35}
\tma^0(\tr_2 ,\tr_1)=
\begin{pmatrix}
\frac{\gamma+\tg \tr_1^3 \tr_2^{-3}}{\gamma+\tg} & 0 & \frac{3\gamma\tg(\tr_2^{-3}-\tr_1^{-3})}{\gamma+\tg} & \frac{3\tal\tg\tr_2^{-3}}{\tc(1+\tg)}\\
\frac{\tal}{\tc(1+\tg)}&0&-\frac{3\tal\tg\tr_1^{-3}}{\tc(1+\tg)}&0\\
\frac{\tr_1^3-\tr_2^3}{3(\gamma+\tg)}&0&\frac{\gamma+\tg\tr_2^3\tr_1^{-3}}{\gamma+\tg} & \frac{\tal}{\tc(1+\tg)}\\
0&0&0&0
\end{pmatrix}\,,
\end{equation}

\begin{equation}\label{b36}
\tma^-(\tr_2 ,\tr_1)=
\begin{pmatrix}
\frac{3\tal^2\tg(1+\tr_2)(\tr_1-1)}{2 \tc(1+\tg)^2 \tr_2^3} & \frac{3\tal\tg(1+\tr_2)(1-\tr_1)}{2(1+\tg) \tr_2^3} & 
\frac{9\tal^2\tg^2(1+\tr_2)(1-\tr_1)}{2 \tc(1+\tg)^2 \tr_2^3 \tr_1^3} &
-\frac{3\tal\tg(1+\tr_2)}{2\tc(1+\tg) \tr_2^3 \tr_1} \\
\frac{\tal(1-\tr_1)}{2\tc(1+\tg)\tr_2} & \frac{\tr_1-1}{2 \tr_2} &
\frac{3 \tal \tg (\tr_1-1)}{2 \tc(1+\tg)\tr_2 \tr_1^3 } & 
\frac{1}{2 \tc \tr_2 \tr_1} \\
\frac{\tal^2(1+\tr_2)(\tr_1-1)}{2 \tc(1+\tg)^2} & \frac{\tal(1+\tr_2)(1-\tr_1)}{2(1+\tg)} & 
\frac{3\tal^2\tg(1+\tr_2)(1-\tr_1)}{2 \tc(1+\tg)^2 \tr_1^3} &
-\frac{\tal(1+\tr_2)}{2\tc(1+\tg)\tr_1} \\
\frac{\tal(1+\tr_2)(1-\tr_1)}{2(1+\tg)} & \frac{\tc(1+\tr_2)(\tr_1-1)}{2} &
\frac{3\tal \tg (1+\tr_2)(\tr_1-1)}{2 (1+\tg)\tr_1^3} & 
\frac{1+\tr_2}{2\tr_1} 
\end{pmatrix}\,,
\end{equation}
and

\begin{equation}\label{b37}
\tma^+(\tr_2 ,\tr_1)=
\begin{pmatrix}
\frac{3\tal^2\tg(1-\tr_2)(1+\tr_1)}{2 \tc(1+\tg)^2 \tr_2^3} & \frac{3\tal\tg(\tr_2-1)(1+\tr_1)}{2(1+\tg) \tr_2^3} & 
\frac{9\tal^2\tg^2(\tr_2-1)(1+\tr_1)}{2 \tc(1+\tg)^2 \tr_2^3 \tr_1^3} &
\frac{3\tal\tg(1-\tr_2)}{2\tc(1+\tg) \tr_2^3 \tr_1} \\
-\frac{\tal(1+\tr_1)}{2\tc(1+\tg)\tr_2} & \frac{1+\tr_1}{2 \tr_2} &
\frac{3 \tal \tg (1+\tr_1)}{2 \tc (1+\tg)\tr_2 \tr_1^3} & 
-\frac{1}{2 \tc \tr_2 \tr_1} \\
\frac{\tal^2(1-\tr_2)(1+\tr_1)}{2 \tc(1+\tg)^2} & \frac{\tal(\tr_2-1)(1+\tr_1)}{2(1+\tg)} & 
\frac{3\tal^2\tg(\tr_2-1)(1+\tr_1)}{2 \tc(1+\tg)^2 \tr_1^3} &
\frac{\tal(1-\tr_2)}{2\tc(1+\tg)\tr_1} \\
\frac{\tal (\tr_2-1)(1+\tr_1)}{2(1+\tg)} & \frac{\tc(1-\tr_2)(1+\tr_1)}{2} &
\frac{3\tal \tg (1-\tr_2)(1+\tr_1)}{2 (1+\tg)\tr_1^3} & 
\frac{\tr_2-1}{2\tr_1}
\end{pmatrix}
\end{equation}

Although the expression for the transfer matrix is cumbersome, $\tma$ has a number of simple mathematical features. First, explicit calculation shows that $\det(\mma(\tr))=2\tc(1+\tg)$ independent of radius. This implies that

\begin{equation}\label{dett}
\det(\tma)
\,=\,1\,.
\end{equation}
Moreover, the definition of $\tma$ immediately implies that $\tma(\tr,\tr)=\textbf{E}$ where  $\textbf{E}$ is the unit matrix, and that the inverse is given by

\begin{equation}\label{b38}
 \tma^{-1}(\tr_2,\tr_1)=\tma(\tr_1,\tr_2).
\end{equation}
More generally, by its definition the transfer matrix clearly obeys the functional equation

\begin{equation}\label{b39}
 \tma(\tr_3,\tr_1)=\tma(\tr_3,\tr_2) \tma(\tr_2,\tr_1)\,.
\end{equation}
This gives rise to a number of relations between $\tma^0$, $\tma^-$ and $\tma^+$:

\begin{eqnarray}\label{b40}
 \tma^0(\tr_3,\tr_1) & = &\tma^0(\tr_3,\tr_2) \tma^0(\tr_2,\tr_1)\,, \nonumber \\
 \tma^-(\tr_3,\tr_1) & = & \tma^-(\tr_3,\tr_2) \tma^-(\tr_2,\tr_1) \,,\\
 \nonumber
 \tma^+(\tr_3,\tr_1)& = &\tma^+(\tr_3,\tr_2) \tma^+(\tr_2,\tr_1)\,,
\end{eqnarray}
and

\begin{eqnarray}\label{b41}
\tma^0(\tr_3,\tr_2) \tma^+(\tr_2,\tr_1)= 0 & , &\tma^0(\tr_3,\tr_2) \tma^-(\tr_2,\tr_1) = 0 \,,\nonumber\\
\tma^+(\tr_3,\tr_2) \tma^0(\tr_2,\tr_1)= 0 & , &\tma^-(\tr_3,\tr_2) \tma^0(\tr_2,\tr_1) = 0 \,,\\
\nonumber
\tma^+(\tr_3,\tr_2) \tma^-(\tr_2,\tr_1)= 0 & , &\tma^-(\tr_3,\tr_2) \tma^+(\tr_2,\tr_1) = 0 \,.
\end{eqnarray}

The dimensionless transfer matrix $\tma(\tr_2,\tr_1)$ given by Eq. ($\ref{b33}$) is related to the dimensionless response matrix $\gmma(\tr_2,\tr_1)$ defined by the relation

\begin{equation}\label{b411}
\begin{pmatrix}
\dtpr(\tr_1)\\ 
\dtt(\tr_1)\\
\dtpr(\tr_2)\\ 
\dtt(\tr_2)
\end{pmatrix}\,=\,
\gmma(\tr_2,\tr_1)
\begin{pmatrix}
\dtV(\tr_1)\\
\dtQ(\tr_1)\\
-\dtV(\tr_2)\\
-\dtQ(\tr_2)
\end{pmatrix}\,.
\end{equation}
If $\tma$ and $\gmma$ are split into four blocks of ($2 \times 2$)-matrices as follows

\begin{equation}\label{b42}
 \tma=
 \begin{pmatrix}
  \tma_1 & \tma_2 \\ \tma_3 & \tma_4
 \end{pmatrix}\,,
\end{equation}
and

\begin{equation}\label{b421}
 \gmma=
 \begin{pmatrix}
  \gmma_1 & \gmma_2 \\ \gmma_3 & \gmma_4
 \end{pmatrix}\,,
\end{equation}
one finds

\begin{eqnarray}\label{b422}
\gmma_1=-\tma_3^{-1} \tma_4 & , & \gmma_2=-\tma_3^{-1}\,, \\
\gmma_3=-\tma_1\tma_3^{-1} \tma_4+\tma_2 & , & \gmma_4=-\tma_1\tma_3^{-1}\,.
\end{eqnarray}
Calculating $\gmma$ from $\tma$ gives

\begin{eqnarray}
\tilde \Gamma_{11}(\tr_2,\tr_1)&=&\frac{3}{\tr_2^3-\tr_1^3} \left(\gamma+\tg (\frac{\tr_2}{\tr_1})^3\right)\\
\tilde \Gamma_{12}(\tr_2,\tr_1) &=&\frac{3\tal \gamma_l}{\tc (\tr_2^3-\tr_1^3)} \\
\tilde \Gamma_{13}(\tr_2,\tr_1) &=&\frac{3(\gamma+\tg)}{\tr_2^3-\tr_1^3} \\
\tilde \Gamma_{22}(\tr_2,\tr_1) &=&\frac{3(\gamma_l-1)}{\tc(\tr_2^3-\tr_1^3)}+\frac{\tr_2\cosh(\tr_2-\tr_1)-\sinh(\tr_2-\tr_1)}{\tc \tr_1[(\tr_2-\tr_1)\cosh(\tr_2-\tr_1)+(\tr_1\tr_2-1)\sinh(\tr_2-\tr_1)]} \\
\tilde \Gamma_{24}(\tr_2,\tr_1) &=&\frac{3(\gamma_l-1)}{\tc(\tr_2^3-\tr_1^3)}+\frac{1}{\tc [(\tr_2-\tr_1)\cosh(\tr_2-\tr_1)+(\tr_1\tr_2-1)\sinh(\tr_2-\tr_1)]}
\end{eqnarray}
All off-diagonal elements of the four block matrices become identical, i.e.,

\begin{equation}
\tilde \Gamma_{14}= \tilde \Gamma_{21}=\tilde\Gamma_{23}=\tilde \Gamma_{32} =\tilde \Gamma_{34}=\tilde \Gamma_{41}=\tilde \Gamma_{43}=\tilde \Gamma_{12}
\end{equation}
Furthermore,

\begin{eqnarray}
\tilde \Gamma_{31}(\tr_2,\tr_1)&= -\tilde \Gamma_{13}(\tr_1,\tr_2)&=\tilde \Gamma_{13}(\tr_2,\tr_1)\,, \\
\tilde \Gamma_{33}(\tr_2,\tr_1)&= -\tilde \Gamma_{11}(\tr_1,\tr_2) & \,,\\
\tilde \Gamma_{42}(\tr_2,\tr_1)&= -\tilde \Gamma_{24}(\tr_1,\tr_2) &=\tilde \Gamma_{24}(\tr_2,\tr_1)\,, \\
\tilde \Gamma_{44}(\tr_2,\tr_1)&= -\tilde \Gamma_{22}(\tr_1,\tr_2) & \,.\\
\end{eqnarray}

The response matrix $\gmma$ is symmetric, i.e., $\gmma_1=\gmma_1'$, $\gmma_2=\gmma_3'$ and $\gmma_4=\gmma_4'$. These Onsager relations follow also from general arguments (the fluctuation-dissipation theorem). Interestingly, $\gmma$ has an even higher symmetry:

\begin{eqnarray}
 \gmma_1(\tr_2,\tr_1)=-\gmma_4(\tr_1,\tr_2) & , & \gmma_2(\tr_2,\tr_1)=-\gmma_3(\tr_1,\tr_2)\,.
\end{eqnarray}
These relations follow from Eq. (\ref{b38}) and the connection between $\gmma$ and $\tma$. On the other hand, the symmetry relation $\gmma=\gmma'$ implies

\begin{eqnarray}\label{b43}
 \tma_4 \tma_3'=\tma_3 \tma_4'  \, , &  \tma_3' \tma_1=\tma_1' \tma_3\ \, , & \textbf{E}=\tma_1' \tma_4-\tma_3' \tma_2\, ,
\end{eqnarray}
and that

\begin{eqnarray}\label{b45}
 \tma_1 \tma_2'=\tma_2 \tma_1' \, , &  \tma_4' \tma_2=\tma_2' \tma_4\,, & \textbf{E}=\tma_1 \tma_4'-\tma_2 \tma_3'\,.
\end{eqnarray}

\section{Some cases of experimental relevance}

\subsection{A massive sphere}

In this first application (Fig. 2) of the formalism we inquire into how a solid sphere responds to a compression $-\delta V$ and a heat supply $-\delta Q$ \cite{footnote} applied at radius $r_2$ (the transferred heat is positive when $\delta Q<0$ because $\delta Q$ refers to the heat flow in the positive radial direction). This is calculated from the transfer matrix by putting $r_1=0$. If $r_2 \ll | l_D|$, we expect the response matrix $\mathbf{G}_0$ to be given by the constitutive equations (5), (7), and (2) generalized to frequency-dependent coefficients (where all variables refer to the outer radius $r_2$):

\begin{equation}\label{ca1}
\begin{pmatrix}
\delta p_r\\
\delta T
\end{pmatrix}\,=\,
\mathbf{G}_0
\begin{pmatrix}
-\delta V\\ 
-\delta Q\\
\end{pmatrix}\,=\,
\frac{1}{V}\begin{pmatrix}
K_S & \frac{1}{T_0\alpha_S}\\
\frac{1}{ \alpha_S} & \frac{1}{c_V}
\end{pmatrix}
\begin{pmatrix}
-\delta V\\ 
-\delta Q\\
\end{pmatrix}\,.
\end{equation}
In general, the relation is

\begin{equation}\label{ca2}
\begin{pmatrix}
\dtpr\\
\dtt
\end{pmatrix}\,=\,
\mathbf{\tilde G}
\begin{pmatrix}
-\delta\tilde V\\ 
-\delta\tilde Q\\
\end{pmatrix}\,
\end{equation}
with

\begin{equation}\label{ca2a}
\mathbf{\tilde G}=-\tma_1(\tr_2,0) \tma_3^{-1}(\tr_2,0)=\gmma_4(\tr_2,0)\,.
\end{equation}
This follows from Eqs. (\ref{b33}) and (\ref{b42}) via the boundary conditions $\dtV(\tr_1=0)=0$ and $\dtQ(\tr_1=0)=0$. Returning to dimensional variables, Eqs. (\ref{ca2}) and (\ref{ca2a}) yields

\begin{equation}\label{ca3}
\begin{pmatrix}
\delta p_r\\
\delta T
\end{pmatrix}\,=\,
\frac{1}{V}\begin{pmatrix}
K_S & \frac{1}{T_0 \alpha_S}\\
\frac{1}{\alpha_S} & \frac{1}{c_V} f_D
\end{pmatrix}
\begin{pmatrix}
-\delta V\\ 
-\delta Q\\
\end{pmatrix},
\end{equation}
where $f_D$ is the function of frequency defined by

\begin{equation}
f_D
\,=\,1+\frac{1}{\gamma_l}\left(\frac{1}{3}\frac{x^2 \sinh(x)}{x \cosh(x)-\sinh(x)}-1 \right)\,,\, \, x=r_2/l_D\,.
\end{equation}
The frequency dependence of $f_D$ derives primarily from that of $l_D$. Note that $f_D \rightarrow 1$ for $\omega \rightarrow 0$ and $f_D \rightarrow \infty$ for $\omega \rightarrow \infty$. Asymptotically, one has $f_D =  r_2/(3 \gamma_l l_D)$ for $\omega \rightarrow \infty$. The function $f_D$ describes how heat diffusion affects the measurement of the thermal and mechanical properties of a massive sphere that can only, of course, be accessed at the surface.

\begin{figure}
\includegraphics[width=8cm]{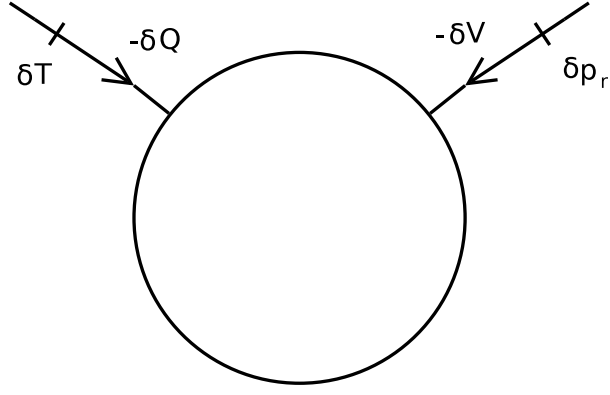} 
\caption{Compression $-\delta V$ and heat input $- \delta Q$ at the surface of a sphere give rise to changes $\delta p_r$ and $\delta T$ in radial pressure and temperature, respectively, at the surface. For a finite sphere the response (Eq. (\ref{ca3})) is given by the constitutive properties $K_S$, $\alpha_S$, and $c_V$, as well as by heat diffusion. Eight kinds of input/output relations are discussed in the text.}
\label{fig:toportsph}
\end{figure}

Below, several different thermoviscoelastic experiments on a massive sphere are considered. In principle there are $24$ such experiments \cite{ell07}, corresponding to the 24 coefficients discussed at the beginning of Sec. II: One may choose any of the four variables $\delta p_r$, $\delta T$, $-\delta V$, or $-\delta Q$ at the outer radius $r_2$ as the controlled input, any of the three remaining variables as the measured output, and any of the last two to be fixed. Since there are only three independent functions of frequency in the matrix of Eq. (\ref{ca3}), it is superfluous to discuss all these 24 experiments. It does make sense, however, to discuss more than just three cases, because the experimental challenges may vary from case to case. 

We discuss eight cases, corresponding in the low-frequency limit to the eight frequency-dependent thermodynamic response functions $\kappa_T, \kappa_S, c_V, c_p, \alpha_p, \alpha_S, \beta_V$, and $\beta_S$. These cases are detailed below where, occasionally, relations from the Appendix are utilized.

\paragraph{Compression with isothermal boundary:}

\begin{equation}
-\left(\frac{\delta p_r}{\delta V}\right)_T
\,=\,G_{11}-\frac{G_{12}G_{21}}{G_{22}}
\,=\,\frac{K_S}{V}\left(1-\frac{1}{f_D}(1-\frac{1}{\gamma})\right)\,.
\end{equation}
At low frequencies this converges to $K_T/V$. This result shows how one is limited upwards in frequency when attempting to do isothermal bulk modulus measurements. Because $f_D\rightarrow\infty$ at large frequencies, the equation also describes the transition to the adiabatic bulk modulus above the characteristic heat-diffusion frequency $\omega_D$ defined by $\omega_D\equiv D/r_2^2$, where $D$ is the heat diffusion constant and $r_2$ the sample size.

The bulk modulus can be measured in the frequency range, $1\,{\rm Hz}-50\,{\rm kHz}$ by the so-called piezoelectric bulk modulus gauge (PBG) \cite{chr94}. This is a piezoelectric ceramic hollow sphere that may be filled with liquid. The ceramic shell has electrodes on the inside and outside and thus constitutes an electrical capacitor. Due to the piezoelectric effect the frequency-dependent capacitance - that can readily be measured - depends on the bulk modulus of the liquid which can be found after a calibration of the PBG. This device has a radius of $10 \,{\rm mm}$ and since typical liquid heat diffusivities are of order $0.1$ ${\rm mm^2/s}$, the characteristic heat diffusion frequency becomes $10^{-3}\,{\rm s}^{-1}$. Thus experiments performed with the PBG above $0.1 \,{\rm Hz}$ can safely be said to be adiabatic, despite the fact that no special measures are taken to make the boundary conditions adiabatic.

\paragraph{Adiabatic compression:}

\begin{equation}
-\left(\frac{\delta p_r}{\delta V}\right)_Q= G_{11}=\frac{K_S}{V}\,.
\end{equation}
Ironically, this boundary condition is difficult to achieve experimentally, whereas the isothermal experiment gives $K_S$ at most frequencies. Thus $K_S$ is easier to measure than $K_T$ \cite{chr94}.

\paragraph{Isochoric heating:}

\begin{equation}
\left(\frac{\delta T}{-\delta Q}\right)_V=G_{22}=\frac{f_D}{Vc_V}\,.
\end{equation}
The low-frequency limit is $1/(Vc_V)$ giving the isochoric frequency-dependent specific heat. The high-frequency limit, however, 

\begin{equation}
\left(\frac{\delta T}{-\delta Q}\right)_V \,\cong\,\frac{1}{4 \pi r^2}\sqrt{\frac{s}{c_l \lambda}} \qquad \mathrm{for} \qquad \omega \rightarrow \infty\,,
\end{equation} 
involves the longitudinal specific heat. Note that in this limit -- even though the overall volume is constant -- it is $c_l$ that appears, not $c_V$.

\paragraph{Heating into a free mechanical surface:}

\begin{equation}
\left(\frac{\delta T}{-\delta Q}\right)_{p_r}=G_{22}-\frac{G_{21}G_{12}}{G_{11}}=\frac{1}{Vc_p}\Big(1+\gamma(f_D-1)\Big)\,.
\end{equation}
The low-frequency limit is $1/(Vc_p)$ giving the isobaric frequency-dependent specific heat. The high-frequency limit,

\begin{equation}\label{a58}
\left(\frac{\delta T}{-\delta Q}\right)_{p_r} \,\cong\, \frac{1}{4 \pi r^2}\sqrt{\frac{s}{c_l \lambda}} \quad \mathrm{for} \quad \omega \rightarrow \infty\,,
\end{equation}
is identical to the isochoric high-frequency limit. Note that it is the longitudinal specific heat that enters into Eq. (\ref{a58}), not $c_p$. This is similar to the fact that the frequency-dependent specific heat obtained from plane-wave effusivity measurements is not the isobaric specific heat, but the longitudinal \cite{chr07}. In that case nonisotropic stresses could be conceived as arising from the special kind of mechanical boundary conditions needed in order to keep the model of the plane-plate setup one-dimensional. Here we see, however, that nonisotropic stresses may arise in the liquid itself, not necessarily coming from clamping boundaries. This substantiates a conclusion of Ref. \cite{chr07}, namely that it is not possible to probe the isobaric specific heat directly by effusivity measurements. -- Note also that the thermal admittance per unit area is the same as for the planar geometry \cite{chr07}:

\begin{equation}
Y\,\equiv\,
-\frac{j}{\delta T}=-\frac{s}{4 \pi r^2 }\frac{\delta Q}{T}=\sqrt{s c_l \lambda}\,.
\end{equation}
Recently, radial heat effusion from the surface of a spherical cavity inside an infinite medium was shown also to involve the longitudinal specific heat \cite{chr08}.

\paragraph{Expansion by a controlled temperature oscillation at a free surface:}

\begin{equation}\label{temp_exp}
\frac{1}{V}\left(\frac{\delta V}{\delta T}\right)_{p_r}\,=\,
\frac{1}{V}\frac{G_{12}}{G_{12}G_{21}-G_{11}G_{22}}
\,=\,\frac{\alpha_p}{1+\gamma(f_D-1)}\,.
\end{equation}
At low frequencies this approaches $\alpha_p$, whereas its high-frequency asymptotic form is given by

\begin{equation}\label{Y_lign}
\frac{1}{V}\left(\frac{\delta V}{\delta T}\right)_{p_r}\,\cong\,
\frac{3}{\beta_Ssr_2}Y\quad \mathrm{for} \quad \omega \rightarrow \infty\,.
\end{equation}
From Eq. (\ref{ca3}) it follows that 

\begin{equation}\label{Max_gen}
\frac{1}{V}\left(\frac{\delta V}{\delta T}\right)_{p_r}\,=\,
-\frac{1}{VT_0}\left(\frac{\delta Q}{\delta p_r}\right)_{T}\,.
\end{equation}
This is a radial version of the Onsager relation that corresponds to the Maxwell relation Eq. (\ref{iso}). If an experiment is conceived where one measures the heat flux $-\delta Q$ needed to keep temperature constant at the surface while applying a periodically varying radial pressure, one would find the same response function Eq. (\ref{temp_exp}), including the diffusion dependence. Similar radial versions of Onsager relations corresponding to the Maxwell relations Eqs. (\ref{ref_1}), (\ref{ref_2}), and (\ref{ref_3}), hold for the last three response functions:

\paragraph{Radial pressure in response to heating at constant volume:}

\begin{equation}\label{Pres_heat}
\left(\frac{\delta p_r}{-\delta Q}\right)_{V}\,=\,
\frac{1}{VT_0\alpha_S}\,.
\end{equation}
Thus $\alpha_S$ can be measured without interference from heat diffusion. This is not trivial, since the penetration depth $|l_D|$ of the temperature field into the sphere is frequency dependent, and it is the temperature field that creates the pressure variation.

\paragraph{Radial pressure in response to a controlled temperature oscillation at constant volume:}

This case leads to

\begin{equation}\label{Pres_temp}
\left(\frac{\delta p_r}{\delta T}\right)_{V}\,=\,
\frac{\beta_V}{f_D}\,.
\end{equation}
In contrast to case (f), this response function {\it is} diffusion influenced. It approaches $\beta_V$ at low frequencies, whereas

\begin{equation}\label{Y_lign_2}
\frac{1}{V}\left(\frac{\delta p_r}{\delta T}\right)_V\,\cong\,
\frac{3}{\alpha_Ssr_2}Y\quad \mathrm{for} \quad \omega \rightarrow \infty\,.
\end{equation}

\paragraph{Volume expansion in response to heating for a free surface:}


\begin{equation}\label{Vol_heat}
\left(\frac{\delta V}{-\delta Q}\right)_{p_r}\,=\,
\frac{1}{T_0\beta_S}\,.
\end{equation}
As for case (f) we get a simple result that is independent of heat diffusion.

In summary, the important role played by the longitudinal frequency-dependent specific heat is evident. Moreover, there is now an exact description of the transition from the adiabatic to the isothermal regimes of bulk modulus measurements utilizing the PBG \cite{chr94}.

\subsection{The ``thermally thick limit'' $ |l_D| \ll r_2$ when $r_1 \ll r_2$}

We now proceed to discuss the case where both the inner radius is small and the sample is much larger than $l_D$. This section prepares the theoretical basis of ongoing experiments where a small spherical thermistor placed in the center of the PBG makes it possible to simultaneously measure the frequency dependences of $\alpha_S$ , $K_S$ and $c_l$ on the same sample. Supplemented by shear modulus measurements \cite{chr95} this provides a complete set of thermoviscoelastic response functions of a liquid. Below we determine the reduced transfer matrix $\xma$ for a situation where a mechanical boundary condition at $r_1$ and a thermal boundary condition at $r_2$ are stipulated. That is, $\xma$ gives the linear relationship

\begin{equation}\label{cc1}
\begin{pmatrix}
\dtt(\tr_1)\\
\dtQ(\tr_1)
\end{pmatrix}\,=\,
\xma
\begin{pmatrix}
\dtpr(\tr_2)\\ 
\dtV(\tr_2)\\
\end{pmatrix}\,.
\end{equation}
There are four possibilities for $\xma$, denoted below by $\ama$, $\bma$, $\cma$, and $\dma$, depending on the different boundary conditions:

\begin{eqnarray}\nonumber
 \xma \,=\,\ama & \mathrm{for} & \dtV(\tr_1)=0 , \: \dtQ(\tr_2)=0 \,,\\\nonumber
 \xma \,=\,\bma  & \mathrm{for} & \dtV(\tr_1)=0 , \: \dtt(\tr_2)=0 \,,\\\nonumber
 \xma \,=\,\cma & \mathrm{for} & \dtpr(\tr_1)=0 , \: \dtQ(\tr_2)=0 \,,\\\nonumber
 \xma \,=\,\dma & \mathrm{for} & \dtpr(\tr_1)=0 , \: \dtt(\tr_2)=0\,. \\\nonumber\,.
\end{eqnarray}
If $\pma$ is the inverse of $\tma(\tr_2,\tr_1)$, i.e., $\pma = \tma(\tr_1,\tr_2)$, one finds

\begin{eqnarray}\label{cc2}
\ama &=&\frac{1}{\tP_{32}}
 \begin{pmatrix}
  \tP_{21}\tP_{32}-\tP_{22}\tP_{31} & \tP_{23}\tP_{32}-\tP_{22}\tP_{33} \\
  \tP_{41}\tP_{32}-\tP_{42}\tP_{31} & \tP_{43}\tP_{32}-\tP_{42}\tP_{33}  
 \end{pmatrix}\,, \\
\bma &=&\frac{1}{\tP_{34}}
 \begin{pmatrix}
  \tP_{21}\tP_{34}-\tP_{24}\tP_{31} & \tP_{23}\tP_{34}-\tP_{24}\tP_{33} \\
  \tP_{41}\tP_{34}-\tP_{44}\tP_{31} & \tP_{43}\tP_{34}-\tP_{44}\tP_{33}  
 \end{pmatrix}\,, \\
\cma &=&\frac{1}{\tP_{12}}
 \begin{pmatrix}
  \tP_{21}\tP_{12}-\tP_{22}\tP_{11} & \tP_{23}\tP_{12}-\tP_{22}\tP_{13} \\
  \tP_{41}\tP_{12}-\tP_{42}\tP_{11} & \tP_{43}\tP_{12}-\tP_{42}\tP_{13}  
 \end{pmatrix}\,, \\
\dma &=&\frac{1}{\tP_{14}}
 \begin{pmatrix}
  \tP_{21}\tP_{14}-\tP_{24}\tP_{11} & \tP_{23}\tP_{14}-\tP_{24}\tP_{13} \\
  \tP_{41}\tP_{14}-\tP_{44}\tP_{11} & \tP_{43}\tP_{14}-\tP_{44}\tP_{13}  
 \end{pmatrix}\,.
\end{eqnarray}
These expressions imply \cite{deten_note} that 

\begin{equation}\label{deteret}
\det(\ama)\,=\,\det(\bma)\,=\,\det(\cma)\,=\,\det(\dma)\,=\,1\,.
\end{equation}
The explicit expressions for $\ama$, $\bma$, $\cma$, and $\dma$ are rather involved. We give only the components of the simplest one, $\ama$:

\begin{eqnarray}\label{cc3}
\tA_{11} =&\frac{(\tr_1^3+3(\gamma_l-1)(\tr_1^2-\tr_1)-\tr_2^3)(\tr_2+1)\exp(-(\tr_2-\tr_1))+(\tr_1^3-3(\gamma_l-1)(\tr_1^2+\tr_1)-\tr_2^3)(\tr_2-1)\exp(\tr_2-\tr_1)}{3 \tal \tr_1 \gamma_l((\tr_1-1)(\tr_2+1)\exp(-(\tr_2-\tr_1))+(1-\tr_2)(\tr_1+1)\exp(\tr_2-\tr_1))}\,, \\ \nonumber
\tA_{12} =&-\frac{(\tg(\tr_1^3+(\gamma_l-1)(3\tr_1^2-3\tr_1+\tr_2^3))+\gamma_l\tr_2^3)(\tr_2+1)\exp(-(\tr_2-\tr_1))+(\tg(\tr_1^3-(\gamma_l-1)(3\tr_1^2+3\tr_1-\tr_2^3))+\gamma_l\tr_2^3)(\tr_2-1)\exp(\tr_2-\tr_1)}{\tr_2^3 \tal \tr_1 \gamma_l((\tr_1-1)(\tr_2+1)\exp(-(\tr_2-\tr_1))+(1-\tr_2)(\tr_1+1)\exp(\tr_2-\tr_1))}\,, \\ \nonumber
\tA_{21}=&\frac{\tc(\tr_2^3-\tr_1^3)}{3\tal\gamma_l}\,, \\ \nonumber
\tA_{22}=&\frac{\tc(\tg\tr_1^3+\gamma\tr_2^3)}{\tal\gamma_l\tr_2^3}\,.
\end{eqnarray}
The transfer matrix $\ama$ becomes much simpler when $r_1 \ll r_2$ and $ |l_D| \ll r_2$. In terms of the scaled variables we seek the limits $\tr_1/\tr_2 \rightarrow 0$ and $|\tr_2| \rightarrow \infty$. In these limits one finds from Eq. (\ref{cc3}) that $\ama \rightarrow \thma$ where

\begin{equation}\label{cc4}
\thma \,=\,
\begin{pmatrix}
\frac{\tr_2^3}{3 \gamma_l \tal \tr_1(1+\tr_1)} & \frac{\gamma}{\gamma_l \tal \tr_1(1+\tr_1)}\\
\frac{\tc \tr_2^3}{3\tal \gamma_l} & \frac{\tc\gamma}{\tal \gamma_l} 
\end{pmatrix}\,.
\end{equation}
By explicit calculation one finds that $\bma$, $\cma$, as well as $\dma$, all converge to $\thma$ in the same limits. That is, in these limits the linear connection between the thermal response at the inner radius and the mechanical stimulus at the outer radius is independent of the mechanical boundary condition at the inner radius or the thermal boundary condition at the outer radius. If one inverts $\ama$, $\bma$, $\cma$, and $\dma$ and go to the same limits, the inverse matrices all converge to the matrix $\psima$ given by

\begin{equation}\label{cc5}
\psima \,=\,
\begin{pmatrix}
\frac{\tc\gamma}{\tal \gamma_l}  & -\frac{\gamma}{\gamma_l \tal \tr_1(1+\tr_1)}\\
-\frac{\tc \tr_2^3}{3\tal \gamma_l} & \frac{\tr_2^3}{3 \tal \gamma_l \tr_1(1+\tr_1)}
\end{pmatrix}\,.
\end{equation}

In order to make the above results more transparent we return to dimensional variables. Thus introducing the sphere volume

\begin{equation}\label{cc7}
V_2=\frac{4 \pi}{3}r_2^3,
\end{equation}
and the quantity $Z_{\textnormal{th}}$ to be identified below with a thermal impedance,
\begin{equation}\label{cc8}
Z_{\textnormal{th}}=\frac{1}{4 \pi \lambda r_1 \left(1+ \sqrt{sr_1^2 c_l/\lambda}\right)}\,,
\end{equation}
Eq.  (\ref{cc4}) becomes

\begin{equation}\label{cc9}
\begin{pmatrix}
\dt(r_1)\\
\delta Q(r_1)
\end{pmatrix}
\,=\,
\begin{pmatrix}
s Z_{\textnormal{th}} V_2 T_0 \alpha_S  & s Z_{\textnormal{th}} T_0 \beta_S \\
V_2 T_0 \alpha_S & T_0 \beta_S 
\end{pmatrix}
\begin{pmatrix}
\delta p_r(r_2)\\ 
\delta V(r_2)\\
\end{pmatrix}\,,
\end{equation}
and Eq. (\ref{cc5}) becomes

\begin{equation}\label{cc10}
\begin{pmatrix}
\delta p_r(r_2)\\ 
\delta V(r_2)\\
\end{pmatrix}
\,=\,
\begin{pmatrix}
T_0 \beta_S & -s Z_{\textnormal{th}} T_0 \beta_S \\
-V_2 T_0 \alpha_S & s Z_{\textnormal{th}} V_2 T_0 \alpha_S  
\end{pmatrix}
\begin{pmatrix}
\dt(r_1)\\
\delta Q(r_1)
\end{pmatrix}\,.
\end{equation}
Three results may be inferred: 1) The adiabatic frequency-dependent bulk modulus is given by the normal-stress response to a compression at the outer surface independent of the other three boundary conditions:

\begin{equation}
-V_2\left (\frac{\delta p_r}{\delta V} \right)(r_2)
\,=\,\frac{\beta_S}{\alpha_S}=K_s\,.
\end{equation}
2) The temperature response on the inner surface to a thermal current $s\delta Q$ is the thermal impedance involving the longitudinal specific heat. This again holds independent of the other three boundary conditions:

\begin{equation}
\left (\frac{\dt}{s \delta Q} \right)(r_1)
\,=\,Z_{\textnormal{th}}\,.
\end{equation}
3) The relation between the heat displacement at radius $r_1$ and the volume and negative normal-stress variations at radius $r_2$ does not involve any ``delay'' caused by thermal diffusion:

\begin{equation}
\delta Q(r_1)
\,=\,T_0 V_2 \alpha_S \  \delta p_r(r_2)+T_0 \beta_S \ \delta V(r_2)\,.
\end{equation}

\subsection{Mechanical boundary conditions}

If we control the boundary conditions solely via isobaric or isochoric constraints, there is a thermal transfer matrix $\yma$ connecting heat and temperature variations at the inner and outer radia:

\begin{equation}\label{ythdef}
\begin{pmatrix}
\dtt(\tr_2)\\
\dtq(\tr_2)
\end{pmatrix}\,=\,
\yma(\tr_2,\tr_1)
\begin{pmatrix}
\dtt(\tr_1)\\ 
\dtq(\tr_1)\\
\end{pmatrix}\,.
\end{equation}
Depending on boundary condition we define \cite{y_footnote}

\begin{eqnarray}\label{y_cases}
\yma(\tr_2,\tr_1)\,&=&\,\hma(\tr_2,\tr_1)\,\,\,\,\,{\rm for} \,\,\,\ \dtV(\tr_1)=0\,\,{\rm and}\,\,\dtV(\tr_2)=0\,,\\
\yma(\tr_2,\tr_1)\,&=&\,\kma(\tr_2,\tr_1)\,\,\,\,\, {\rm for} \,\,\,\ \dtV(\tr_1)=0\,\,{\rm and}\,\,\dtpr(\tr_2)=0\,,\\
\yma(\tr_2,\tr_1)\,&=&\,\lma(\tr_2,\tr_1)\,\,\,\,\,\, {\rm for} \,\,\,\ \dtpr(\tr_1)=0\,\,{\rm and}\,\,\dtV(\tr_2)=0\,,\\
\yma(\tr_2,\tr_1)\,&=&\,\nma(\tr_2,\tr_1)\,\,\,\,\,{\rm for} \,\,\,\  \dtpr(\tr_1)=0\,\,{\rm and}\,\,\dtpr(\tr_2)=0\,.
\end{eqnarray}
It can be shown \cite{deten_note} that in all four cases one has

\begin{equation}\label{dety}
\det(\yma)\,=\,1\,.
\end{equation}
The following identities follow trivially 

\begin{eqnarray}\label{hklmid}
\hma^{-1}(\tr_2,\tr_1)\,&=&\,\hma(\tr_1,\tr_2)\\
\kma^{-1}(\tr_2,\tr_1)\,&=&\,\lma(\tr_1,\tr_2)\\
\nma^{-1}(\tr_2,\tr_1)\,&=&\,\nma(\tr_1,\tr_2)\,.
\end{eqnarray}

\begin{figure}
\includegraphics[width=8cm]{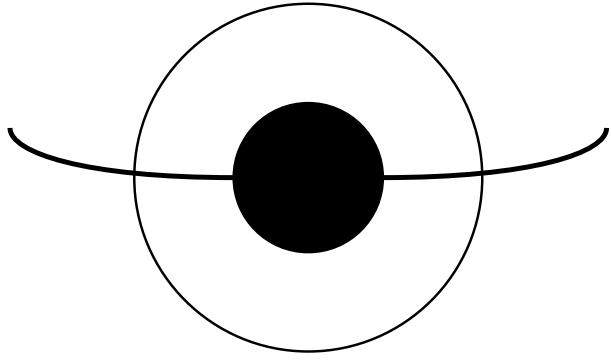}
\caption{AC-calorimetry in spherical geometry \cite{chr98}. A liquid droplet of of diameter $0.7$ mm is placed onto a strongly temperature-dependent resistor of radius $0.3$mm. By the so-called $3\omega$ technique the apparent heat capacity $C_{app}=\delta Q_1/{\delta T_1}$ was found. At low frequencies in a narrow frequency range this gives to a good approximation the isobaric specific heat via Eq. (\ref{C}). At higher frequencies heat diffusion makes the extraction of the frequency-dependent specific heat more involved.}
\label{fig:ntcbead}
\end{figure}

\begin{figure}
\includegraphics[width=12cm]{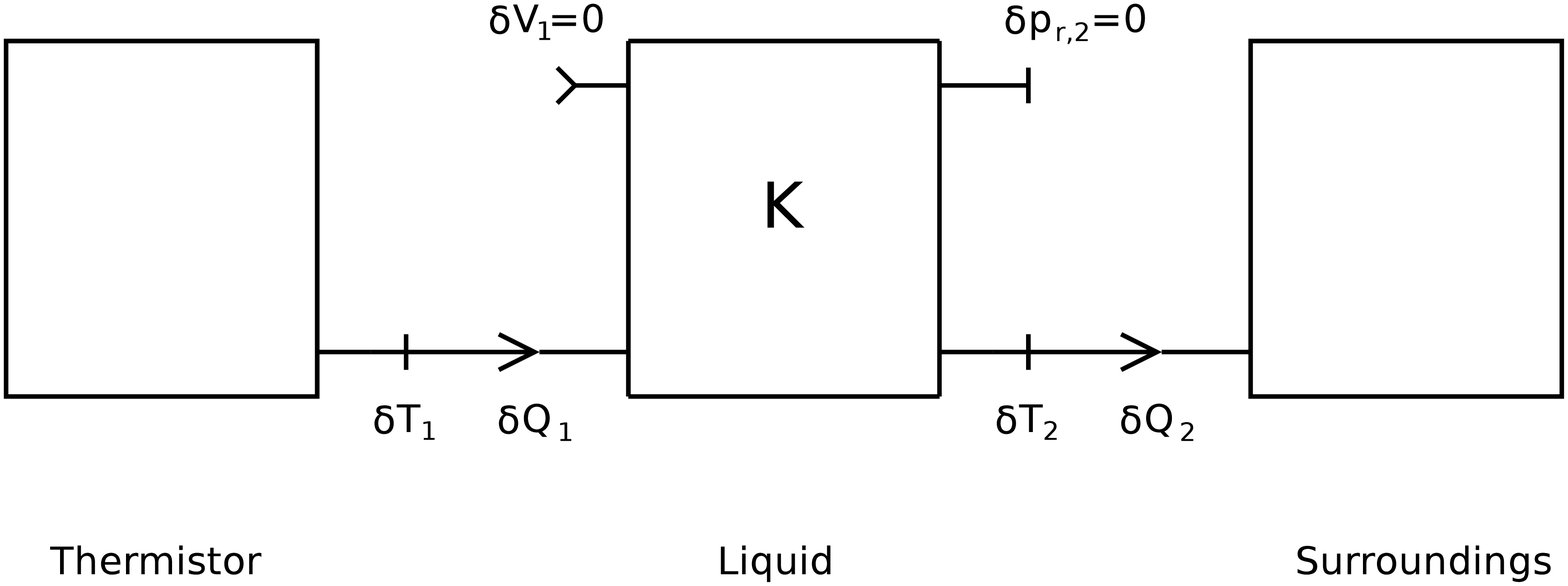}
\caption{Energy bond graph diagram \cite{pvc} of the physical interactions between the thermistor, liquid, and thermal leak to the cryostat, modeling the setup of Fig. 3. The thermistor acts as both heat source and thermometer. The liquid is described by the transfer matrix $K$ for the case with mechanical clamping at the thermistor and a free outer surface.}
\label{fig:ntcbeadeb}
\end{figure}

Motivated by an experimental setup studying a liquid drop on a thermistor bead \cite{chr98} we calculate $\kma$ explicitly (see Figs. \ref{fig:ntcbead} and \ref{fig:ntcbeadeb}). In analogy to the calculation leading to Eq. (\ref{cc2}) we find in terms of the matrix elements of $\tma$ (where all $\tma$ matrix elements are evaluated at $(\tr_2,\tr_1)$)

\begin{equation}\label{kcalc}
\kma(\tr_2,\tr_1)\,=\,
\frac{1}{\ttemp_{11}}
 \begin{pmatrix}
  {\ttemp_{22}\ttemp_{11}-\ttemp_{21}\ttemp_{12}} & {\ttemp_{24}\ttemp_{11}-\ttemp_{21}\ttemp_{14}} \\
  {\ttemp_{42}\ttemp_{11}-\ttemp_{41}\ttemp_{12}} & {\ttemp_{44}\ttemp_{11}-\ttemp_{41}\ttemp_{14}}
 \end{pmatrix} \,.
\end{equation}
Switching to real units, $\kk$ is defined via

\begin{equation}\label{kdef}
\begin{pmatrix}
\delta T(r_2)\\
\delta Q(r_2)
\end{pmatrix}\,=\,
\kk
\begin{pmatrix}
\delta T(r_1)\\ 
\delta Q(r_1)\\
\end{pmatrix}\,,
\end{equation}
and related to $\kma$ by (where $k= 1/l_D$)

\begin{eqnarray}\label{kktrel}
K_{11}(r_2,r_1)\,&=&\,\tilde K_{11}(kr_2,kr_1)\,,\\
K_{12}(r_2,r_1)\,&=&\, \frac{T_0k^3}{4\pi K_T}\, \tilde K_{12}(kr_2,kr_1)\,,\\
K_{21}(r_2,r_1)\,&=&\, \frac{4\pi K_T}{T_0k^3} \,\tilde K_{21}(kr_2,kr_1)\,,\\
K_{22}(r_2,r_1)\,&=&\,\tilde K_{22}(kr_2,kr_1)\,.
\end{eqnarray}
We shall not explicitly give the components for the general case, but limit ourselves to the thermally thin limit $r_2\ll |l_D|$ (i.e., $|kr_2|\ll 1$) where the results simplify considerably. In this limit one finds \cite{chr98} after Taylor expanding the matrix elements of $\tma$ that

\begin{equation}\label{ththin}
\kk\,=\,
\begin{pmatrix}
1 & -Rs\\
-C & 1
\end{pmatrix}
\end{equation}
where, if $V=(4\pi/3)(r_2^3-r_1^3)$ is the volume,

\begin{equation}\label{R}
R\,=\,\frac{1}{4\pi\lambda}\left( \frac{1}{r_1} - \frac{1}{r_2} \right)
\end{equation}
and

\begin{equation}\label{C}
C\,=\,c_V\,V\,\,\frac{K_S(r_2/r_1)^3+\frac{4}{3}G}{K_T(r_2/r_1)^3+\frac{4}{3}G}
\end{equation}
play the role of thermal resistance and capacitance, respectively. These results imply that in the thermally thin limit $c_p$ is measured if $r_2\gg r_1$, whereas $c_l$ is measured if $r_2\cong r_1$. We finally note that the above formulation is easily incorporated into a model taking into account the thermal heat loss to the surroundings \cite{chr98}.

\subsection{No thermomechanical coupling}

As mentioned in the beginning of Sec. III, if the isobaric thermal expansion coefficient $\alpha_p$ is zero, one has $\beta_V=0$ and there is no thermomechanical coupling. In this case, the components $\ttemp_{21}$, $\ttemp_{23}$, $\ttemp_{41}$, and $\ttemp_{43}$ vanish, implying that heat and temperature variations at $\tr_2$ only depend on heat and temperature variations at $\tr_1$. Also, $\ttemp_{12}$, $\ttemp_{14}$, $\ttemp_{32}$, and $\ttemp_{34}$ vanish, implying that pressure and volume variations at $\tr_2$ depend only on pressure and temperature variations at $\tr_1$.  Note that when there is no thermomechanical coupling, all specific heats are identical: 

\begin{equation}\label{cid}
c_p\,=\,c_V\,=\,c_l\,.
\end{equation}
This is often a good approximation for solids, but rarely for liquids.

When there is no thermomechanical coupling, heat diffusion is described by a $2\times 2$ thermal transfer matrix $\tth$ defined \cite{th_footnote} as follows,

\begin{equation}\label{th_diff}
\begin{pmatrix}
\delta T(r_2)\\
\delta Q (r_2)
\end{pmatrix}\,=\,
\tth(r_2,r_1)
\begin{pmatrix}
\delta T(r_1)\\ 
\delta Q(r_1)
\end{pmatrix}\,.
\end{equation}
The components of $\tth$ are found by substituting $\alpha_p=0$ into Eq. (\ref{b34}). The results are as follows (where $k=1/l_D$)

\begin{eqnarray}\label{tth_comp}
T^{\rm th}_{11}
\,&=&\, \frac{r_1}{r_2}\cosh\big(k(r_2-r_1)\big) + \frac{1}{kr_2}\sinh\big(k(r_2-r_1)\big)\,,\\
T^{\rm th}_{12}
\,&=&\, -\frac{s}{4\pi\lambda}\frac{\sinh\big(k(r_2-r_1)\big)}{kr_1r_2}\,,\\
T^{\rm th}_{21}
\,&=&\, \frac{4\pi c_l}{k^3 }\Big[\big(1-k^2r_1r_2\big)  \sinh\big(k(r_2-r_1)\big) - k(r_2-r_1)\cosh\big(k(r_2-r_1)\big)\Big]\,,\\
T^{\rm th}_{22}
\,&=&\, \frac{r_2}{r_1}\cosh\big(k(r_2-r_1)\big) - \frac{1}{kr_1}\sinh\big(k(r_2-r_1)\big)\,.\\
\end{eqnarray}
Interestingly, the same purely thermal $2\times 2$ transfer matrix describes a low-viscosity liquid ($\tg\rightarrow 0$) even when $\alpha_p\neq 0$, if either the inner or the outer surface is free, i.e., if $\delta p_r(r_1)=0$ or if $\delta p_r(r_2)=0$. In these two cases, however, not all three specific heats are identical, only $c_p=c_l$ applies. This is consistent with the above remark regarding the validity of the standard heat diffusion equation (\ref{b7}).

\section{Concluding remarks}

Thermoviscoelastic response functions are notoriously difficult to measure. This paper establishes the theoretical framework necessary for developing experimental methods that utilize spherical symmetry for measuring such response functions. From the complete solution of the problem in the form of the transfer matrix the equations describing any realistic experimental situation may be derived, as exemplified in the last section.

The thermoviscoelastic response functions are important to determine for liquids approaching the glass transition (still in metastable equilibrium above the transition). For such ultraviscous liquids all thermodynamic coefficients become complex and frequency dependent for frequencies in the range of the inverse Maxwell relaxation time. To the best of our knowledge there are yet no reliable measurements of a complete set (i.e., three \cite{mei59,chr82,now86,ell07,bai08}) of thermoviscoelastic response functions for any such liquid. The determination of such complete sets, from which all other thermoviscoelastic response functions are easily calculated, serves the obvious purpose of elucidating the macroscopic dynamics and thermodynamics of ultraviscous liquids. Very recent theoretical developments even further stress the importance of developing reliable methods for measuring thermoviscoelastic response functions. It now appears that the class of van der Waals liquids (possibly supplemented by some liquids forming bulk metallic glasses) have particularly simple properties: Liquids with non-directional chemical bonds exhibit strong correlations between equilibrium pressure and energy fluctuations \cite{bai08,ped08}. For such ``strongly correlating viscous liquids'' it has been shown that there is basically only one independent thermoviscoelastic response function \cite{ell07}. It would be interesting to have this prediction subjected to experimental tests. Moreover, for strongly correlating viscous liquids there are strong indications from computer simulations that thermoviscoelastic measurements can determine the exponent of the so-called density scaling that collapses the relaxation time's pressure and temperature dependence onto a master curve \cite{sch08}.

As regards the above results, it is notable that the longitudinal specific heat $c_l$ plays a dominant role. It is perhaps not surprising that $c_l$ enters repeatedly into the equations describing the one-dimensional case \cite{chr07} -- after all, this is what it was defined to do -- but it is less obvious that $c_l$ also plays a dominant role for the case of spherical symmetry. Note, however, that this result is implicit already in Ref. \cite{lan86}.

\acknowledgments 
We are indebted to our mentors Niels Boye Olsen and Peder Voetmann Christiansen for inspiring to this work and to Jacob Jacobsen for checking the calculations. This work was supported by the Danish National Research Foundation's (DNRF) centre for viscous liquid dynamics ``Glass and Time.''

\appendix*\section{}

If $\zeta_p\equiv c_p/T_0$ and $\zeta_V\equiv c_V/T_0$, several standard thermodynamic relations between the eight linear-response functions $\zeta_p, \zeta_V, \kappa_T, \kappa_S, \alpha_p, \alpha_S, \beta_V, \beta_S$ are summarized below. Via the correspondence principle all relations apply also for the corresponding complex, frequency-dependent linear-response properties. Note that the inverse of the isothermal bulk modulus $K_T$ is the compressibility,  $1/K_T=\kappa_T$, and similarly for the adiabatic bulk modulus, $1/K_S=\kappa_S$ .

\begin{eqnarray}
\zeta_p-\zeta_V =  \dfrac{\alpha_p^2}{\kappa_T}  =  \kappa_T\beta_V^2 =\alpha_p \beta_V\\
\dfrac{1}{\zeta_V}-\dfrac{1}{\zeta_p} =  \dfrac{\kappa_S}{\alpha_S^2}  =  \dfrac{1}{\kappa_S\beta_S^2} =\dfrac{1}{\alpha_S \beta_S}\\
\kappa_T-\kappa_S =  \dfrac{\alpha_p^2}{\zeta_p}  =  \dfrac{\zeta_p}{\beta_S^2} =\dfrac{\alpha_p}{\beta_S}\\
\dfrac{1}{\kappa_S}-\dfrac{1}{\kappa_T} =  \dfrac{\zeta_V}{\alpha_S^2}  =  \dfrac{\beta_V^2}{\zeta_V} = \dfrac{\beta_V}{\alpha_S}\\
\beta_V=\dfrac{\alpha_p}{\kappa_T}=\dfrac{\zeta_V}{\alpha_S}\\
\beta_S=\dfrac{\alpha_S}{\kappa_S}=\dfrac{\zeta_p}{\alpha_p}\\
\gamma=\dfrac{\zeta_p}{\zeta_V}=\dfrac{\kappa_T}{\kappa_S}=1+\dfrac{\alpha_p}{\alpha_S}=\dfrac{1}{1-\dfrac{\beta_V}{\beta_S}}
\end{eqnarray}


\begin{thebibliography}{99}
\bibitem{coupling_note} The coupling between mechanics and thermodynamics appears in the constitutive equation Eq. (\ref{duhamel}), where mechanics and thermodynamics only couple when the thermal expansion is nonzero. The coupling, however, also appears in the equations of motion and via the boundary conditions whenever the shear modulus is comparable to the bulk modulus.
\bibitem{bem} The longitudinal heat capacity appears in Eq. (32.4) of Ref. \cite{lan86} (left-hand side coefficient) as the relevant quantity entering the heat conduction equation for a solid in the case where at infinity the temperature tends to a constant value and there is no deformation.
\bibitem{footnote} Note that $j$ is the heat current density outwards in the radial direction.
\bibitem{deten_note} It is a general property of the energy bond formalism that any transfer matrix expressing the variables of one (or more) energy bond in terms of those of one (or more) other bonds has determinant plus or minus unity (the sign depends on the signs of the energy bond variables, see, e.g., Ref. \cite{pvc}). This generalizes the well-known thermodynamic identity that the determinant of the Jacobian for the two functions $S(p,V)$ and $T(p,V)$ is minus one.
\bibitem{y_footnote} The matrices $\hma$, $\kma$, $\lma$, and $\nma$ do not satisfy the analog of the relation Eq. (\ref{b39}) because they do not convey the information of the mechanical fields on an intermediate surface between the outer surfaces at $r_1$ and $r_2$.
\bibitem{th_footnote} The matrix $\tth$ satisfies the functional equation (\ref{b39}) and it obeys $\det(\tth)=1$ [as well as ${\tth}^{-1}(\tr_1,\tr_2)=\tth(\tr_2,\tr_1)$ that follows from Eq. (\ref{b39})].
\end{thebibliography}
\end{document}